# An introduction to the mathematical modelling of iPSCs

L E Wadkin[1*], S Orozco-Fuentes[1], I Neganova[2], M Lako[3], N G Parker[1] and A Shukurov[1]

[1] School of Mathematics, Statistics and Physics, Newcastle University, UK, NE17RU
[2] Institute of Cytology, RAS St Petersburg, Russia
[3] Bioscience Institute, Newcastle University, UK, NE17RU
[*] Corresponding author: l.e.wadkin@ncl.ac.uk

**Keywords:** mathematical modelling, agent-based modelling, mathematical biology, differential equations, random walks, iPSC reprogramming models, pluripotency, hPSCs, cell migration, population models

**Abstract:** The aim of this chapter is to convey the importance and usefulness of mathematical modelling as a tool to achieve a deeper understanding of stem cell biology. We introduce key mathematical concepts (random walk theory, differential equations and agent-based modelling) which form the basis of current descriptions of induced pluripotent stem cells. We hope to encourage a meaningful dialogue between biologists and mathematicians and highlight the value of such an interdisciplinary approach.

# 1 Introduction

Mathematics is a powerful tool to achieve a deeper understanding of biological systems. The application of mathematics to biology has led to many significant achievements in medicine and epidemiology (predicting the spread of 'mad cow' disease [1, 2] and influenza [3]), evolutionary biology [4] and cellular biology (descriptions of chemotaxis [5] and predicting cancer tumour growth [6]). Similarly, the use of mathematics in stem cell research is advancing current knowledge of underlying behaviours which may be difficult to elucidate experimentally and guiding experimental optimisations and protocol development [7-10]. An excellent introduction to the general subject of mathematical biology can be found in Refs [11] and [12].

Mathematical biology uses established concepts from mathematics and theoretical physics to reveal generic aspects of biological behaviour and identify their deep causes. Mathematical and computational modelling provides a framework for rigorous characterisation, the prediction of observations, and a profound understanding of the underlying natural processes.

The wide applicability of these notions to human pluripotent stem cells (hPSCs), including induced pluripotent stem cells (iPSCs), is evident due to their systematic behaviours. There are many biological properties for which mathematical concepts are pertinent: the idea of pluripotency as a potential and statistical macrostate [13], cell fates as 'steady states' [14, 15], random walks and diffusive migrations of pluripotent stem cells [16], and the emergence of Turing-like spatial patterning [17] are just a few relevant examples. The first mathematical model of stem cells in 1964 used stochastic techniques to capture cell fate decisions [18] and has since been extended to cover gene regulation [19-21], molecular states and the cell cycle [22], and cell population dynamics [23] based on experimental results.

In particular, when mathematical models are rigorously underpinned and validated by experimental observations, the reciprocal benefit for experimentation can be profound: an example is the development of an experimentally trained model of iPSC programming, which led to strategies for marked improvements in reprogramming efficiency [8] (see Section 5).

Arguably the most important obstacle in expanding applications of mathematical modelling of stem cell biology is the natural gap between the knowledge and research approaches of experimentalists and theoreticians. Bridging this gap would lead to significant advances, as physics has demonstrated so spectacularly.

It is, however, easy to understand that the process of narrowing the gap is difficult and requires motivation and devotion on both sides. This chapter aims at presenting mathematical ideas and approaches at a level that does not require more than elementary mathematics. This text cannot replace systematic discussion of mathematical biology [11, 12], but we hope it can help to prepare ground for a fruitful dialogue between experimentalists and mathematicians. The next section discusses in more detail why mathematics is an exciting and meaningful method to develop advances in stem cell biology.

## 1.1 Why mathematics for stem cells?

An interdisciplinary approach combining mathematics and experimentation not only furthers our understanding of the underlying biological system, but also has direct implications for experimental optimisation. Below we list some of the key advantages of this approach, framed by recent advances in stem cell biology.

**1.     Identification of systematic behaviours**

The use of concepts from mathematics and physics allows the identification and verification of systematic behaviours within biological systems. For example, the mathematical analysis of hPSC migration experiments within a random walk framework has led to the quantification and classification of the systematic migration of single and pairs of cells [16, 24, 25], as we discuss in Section 2.

**2.     Identifying universal behaviours**

Not only does mathematics reveal systematic behaviours within a system, it can also highlight those which are universal. These universal behaviours reflect fundamental features of a biological system and are captured by identical or closely related governing equations; in turn, these highlight identical or related biological behaviour. Such models are then easily adaptable with simple parameter changes to different cell lines and experimental conditions. Even simple and well-established mathematical concepts (such as logistic growth or random walks) can be rich and flexible in their specific applications. An example is the universal concept of Brownian motion used to describe the random dispersion of particles of an impressively diverse nature – atoms [26], animals searching for food [27], people in a crowd [28] and, closer to our present subject, stem cell migration [16, 24, 29]. As the starting point for modelling cell migration, it provides a comprehensive, flexible and well understood framework for quantifying ensembles of cells and identifying any deviations from completely random movements and their implications on migration-induced clonality loss [24, 30]. Useful and interesting mathematical extensions include geometric Brownian motion [31] which can provide the addition of a systematic drift term (e.g., due to chemotaxis) and fractional Brownian motion [32] which could capture persistence of cell movements.

**3.     Framework for comparison, testing and predicting**

A quantitative framework provides a clear basis for the comparison of different experiments from which the similarities and differences between biological conditions can be probed with quantitative precision and rigour. Firstly, statistical analysis of experimental data allows the quantification of stem cell behaviour which then informs the development of models [16, 24, 33]. Even at this early stage in the model building process, such qualifications reveal interesting insights for experimental comparisons. For example, the quantitative comparison of the migration of individual hPSCs under different experimental conditions concluded that the addition of a CellTrace™, a labelling dye commonly used to track cell generations, significantly reduces cell motility [16]. Once developed, coherent mathematical models provide a non-invasive prognostic modelling tool for rigorous testing and the prediction of experimental behaviour. For iPSCs, mathematical reprogramming models have led to the suggestion of two mechanisms for reprogramming [22] and identified different modes of reprogramming dynamics [34].

## 4. Guiding experimental decision making

Mathematical models and quantitative frameworks can assist with the objective decision making required in the development of experimental protocols. A simple hPSC spatial colony growth model results in an equation allowing the prediction of the time at which adjacent colonies will first merge due to cell proliferation [7]. This result can guide experimentalists to select their cell seeding densities to optimise the colony clonality. Mathematical modelling is more efficient when in a continual feedback loop with experimental results; model development and refinement are informed by experimental data and the modelling results advise the focus and analysis of the experiments. Computational simulations of mathematical models are also cheap to run compared to laboratory experiments. They provide an efficient cost-effective *in-silico* mode of experimentation.

## 5. Strategic approaches for complex systems

Developing coherent models of stem cells and their colonies remains challenging due to their complex behaviours across multiple scales. There are behaviours on the intra-cellular scale (processes happening within cells, such as the cell cycle and PTF regulation), the cellular or micro-environment (the environmental effects on individual cells, such as cell-cell interactions and cell migration) and the colony scale (collective behaviours throughout colonies such as the spatial patterning of differentiation), as illustrated in Figure 1. Building coherent models of multi-scale systems requires a careful strategic approach. Mathematical techniques allow for a bottom-up strategy; key cell properties are first considered in isolation using the simplest models (minimal models) to identify the building blocks of the complexity. This systematic approach leads to the development of a hierarchy of models suitable for a range of behaviours and applications. These models can then be combined with others and developed to the level of sophistication required to capture the whole system. An informative review discussing multi-scale models from a bioengineering approach and their successes in stem cell biology is given in Ref. [35].

*Figure 1 – The complex multi-scale behaviour of stem cells.*

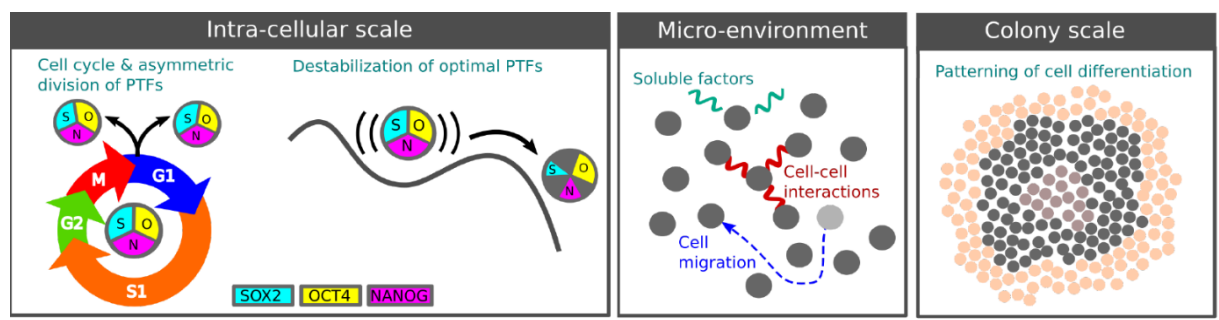

## 1.2 Chapter overview

Although the range of mathematical tools applicable to building the hierarchical model of stem cell behaviour is vast, there are some common fundamental techniques underpinning most current models. These equations and techniques form the basis of more complex mathematical models currently under development for hPSC behaviour and iPSC reprogramming. Each section below explores a different mathematical technique used for the analysis and modelling of stem cell properties.

We first introduce the mathematical concept to make it meaningful to a reader without mathematical background, before discussing some recent examples of the models applied to hPSCs. The aim of this Chapter is to give biological readers an overview of the types of mathematics useful in stem cell research. Although in later sections the mathematics becomes unavoidably more complex, it is presented to give the reader a flavour of the mathematical principles and their applications. We discuss the following concepts and aspects of recent progress in modelling hPSCs:

Section 2: Cell migration as a random walk
> *The theory of random walks is introduced and the migration of individual cells within the random walk framework is discussed.*

Section 3: Differential equations
> *Differential equations are a popular and versatile mathematical tool used to model a wide range of stem cell behaviours. Introductory differential equations are considered before a discussion of differential and stochastic differential models for colony proliferation and cell pluripotency.*

Section 4: Agent-based modelling of colonies
> *The concept of agent-based modelling is explained and a range of agent-based mathematical models for cell migration within colonies and cell proliferation are presented.*

Section 5: iPSC specific models
> *Recent developments in mathematical models specifically for iPSC reprogramming are reviewed and summarised.*

Section 6: Discussion and prospects
> *The next steps in using mathematical modelling to advance the development of iPSC experiments are discussed.*

# 2 Cell migration as a random walk

*Synopsis: This section introduces the theory of random walks (isotropic, biased and correlated) and discusses its successful and informative application to the migration of human pluripotent stem cells.*

Migration is an intrinsic property of many different cell types, including stem cells, and represents a distinct feature of cell behaviour. Although the migration of different cell types may differ in important details, there are fundamental similarities. To begin, the internal mechanics of cell movement can often be ignored to focus on quantifying the properties of migration since they are largely independent of the specific mechanisms of motility. This quantification of cell motion provides a framework for comparison to other cell types and to cells under different experimental conditions. As in many other applications, mathematical approaches provide an opportunity to separate, and then explore, generic and specific aspects of a complex phenomenon.

The classic mathematical description of cell motion is that of a random walk, a sequence of steps in random directions and of random lengths that form the migratory trajectory. Different cell types and different environments are uniquely characterised by the details of the random selection of the direction and distance of movement for each step. Although it is known that cells interact with each other and their environment, models still often take a similar probabilistic approach to simulating cell movement [9, 36]. In this section the mathematical framework for random walks is described, along with its extensions to correlated and biased random walks. Further mathematical details on the use of random walk models in biology can be found in Refs [11, 37] and [38].

## 2.1 Description of random walks
### 2.1.1 Isotropic random walks

Analysing a tortuous, apparently random trajectory of a cell's movement within a random walk framework leads to the identification of parameters which can be used to quantitatively describe cell migration and assist in predictive modelling [16, 24, 25, 29, 39, 40].

Firstly, consider the migration of a cell on a flat surface (a two-dimensional, 2D, system). The simplest is an isotropic random walk (IRW) that has no preferred direction in the cell movement. In this type of model, a continuous motion is approximated by a sequence of steps of a certain length. The shorter is such as step, the more accurate is the approximation. The direction of each step is arbitrary and independent of the earlier movements (this independence is known as the Markovian property). It is natural to expect that cell migration *in-vitro* is isotropic in the absence of large-scale gradients in the environment, and far away from any boundaries.

An idealisation involved in the IRW description is the assumption that a cell moves along a straight line for a short period of time $\tau$, covering a distance $l$, and then changes its direction of migration at random, with each direction having equal probability. Despite its simplicity, the model has been shown to capture a wide range of natural phenomena and is especially relevant to time-lapse imaging of cell migration where frames are taken at discrete times. The time, $\tau$, is an intrinsic property of the migrating cell and characterises the internal motility mechanisms and is unrelated to the frequency of the image recording.

Although each individual cell movement is caused by many influences, the outcome of many such movements develops universal properties sensitive only to a few features of the instantaneous cell behaviour [16, 24, 25]. In particular, if two cells start close to each other, their trajectories will unavoidably diverge and, after a large number of steps, their separations will grow with time $t$ as $\sqrt{t}$. Similarly, the size of a region occupied by a group of cells, each involved in an independent random walk, grows with time as $\sqrt{t}$. This behaviour is observed in many systems including the spread (diffusion) of smell in air, dye in a liquid, heat from a flame, and is known as the diffusive behaviour.

To quantify such movement, consider a migrating cell with a current position $\vec{x}(t)$ and a starting position $\vec{x}_0 = \vec{x}(0)$. In each time step the cell makes a movement in a direction chosen at random. The diffusive nature of this random walk can be quantified by considering the mean square displacement (MSD) of the cell's migratory trajectory. The MSD is a measure of the deviation of a cell from its starting position over time given by $\text{MSD}(t) = \overline{(\vec{x}(t) - \vec{x}_0)^2}$ where the overbar denotes the average taken over a large number of cells. It is a measure of the portion of space explored by the random walker. When the cell's motion is diffusive, the MSD increases linearly with time, $\text{MSD} = 2Dt$, where $D \approx l^2/2\tau$ is the diffusivity, a single summary parameter that characterises an IRW completely at long time or space intervals and fully characterises the cell mobility. Thus, the long-term motion of cells is controlled not by their speed $v = l/\tau$, or distance travelled over a short time interval, but by their combination $D = l^2/2\tau = vl/2$. Depending on the details of cell motion and environments, the MSD may increase slower than $t$ (then the motion is called sub-diffusive) or faster than $t$ (this is super-diffusive behaviour). The variation of the MSD with time for each diffusion type and corresponding trajectories are shown in Figure 2. Note that sub-diffusion can occur due to a combination of displacement and waiting times, as commonly seen in animal movements [37, 41, 42].

*Figure 2 – The MSD with time (top row) for diffusive (left), super-diffusive (middle) and sub-diffusive (right) cell motion and examples of the corresponding cell trajectories (bottom row).*

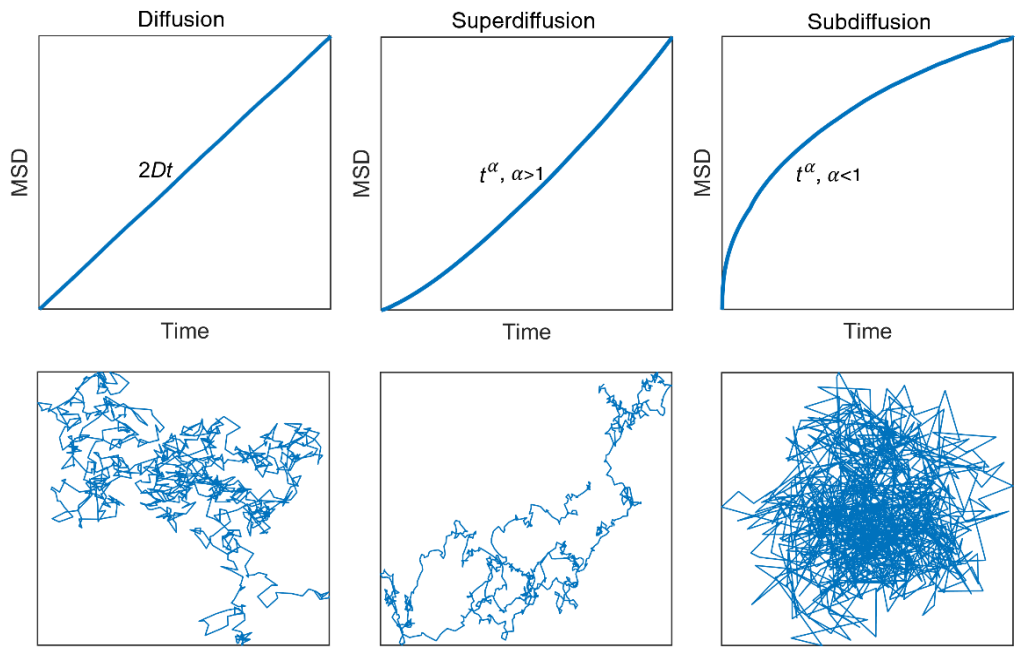

IRWs have been used to successfully describe the motion of hPSCs out of dense colonies (e.g., soon after seeding) [16, 24]. There are many ways to simulate the random walk migration of cells. For computational convenience, the cell positions can be restricted to a discrete lattice or grid, with cells only able to move between fixed positions on the lattice. Some models only allow one cell to occupy any given position on the grid at any time. There are also models in which the cells can move in continuous space. Both types of models, illustrated in Figure 3, are used for individual cell modelling and more complex models of colony kinematics.

*Figure 3 – The random walk migration of cells from a starting position $(x_1, y_1)$ to subsequent positions $(x_i, y_i)$ can be modelled either on a discrete lattice (left) or in continuous space (right). Dashed circles show the cell's previous positions with the filled cell representing its final position.*

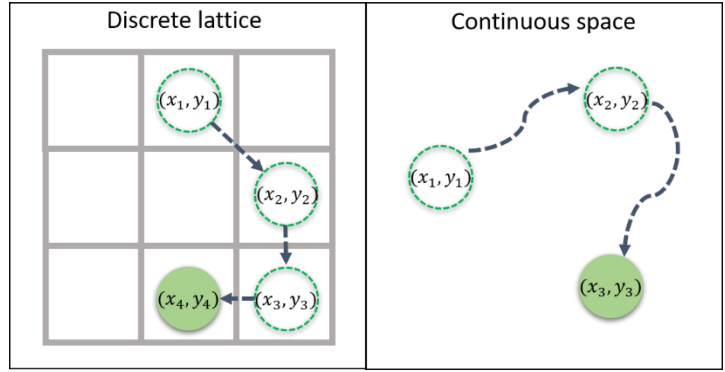

A further useful measure to quantify the overall shape of a cell's trajectory is the directionality (sometimes referred to as the straightness index). The directionality, $\Delta$, of a cell is defined as the ratio between its displacement $r$ (i.e., the shortest distance between the cell's current position and its initial position), and its total distance traversed $d$: $\Delta = r/d$ with $0 \leq \Delta \leq 1$. If the cell moves along a straight path then $r = d$, and the directionality has its maximum value, $\Delta = 1$. If the cell follows a long and tortuous trajectory, then $d$ is much larger than $r$, and the directionality is low, $\Delta \approx 0$. Thus, the directionality quantifies how tangled and convoluted the cell's trajectory is. In an IRW, on average, we have $r = \sqrt{\text{MSD}} = \sqrt{2Dt}$ and $d = vt$, where $v$ is the instantaneous cell speed and $t$ is the time elapsed since the start of the motion. Thus, $\Delta = \sqrt{\tau/t}$, where $\tau$ is the duration of a cell's persistent motion. This quantity is closely related to the tortuosity, similarly characterising the shape of convoluted trajectories commonly used in the analysis of animal movements [43, 44].

When the time scale $\tau$ of the IRW is much shorter than the duration $t$ of an experiment, the discrete nature of the random walk model becomes insignificant. The resulting continuous random walk is described by the diffusion equation which can be considered in addition to, or instead of, simulations of individual random walkers. The diffusion equation is a partial differential equation for the density $n = n(x, y, t)$ of diffusing cells (the number of cells per unit area)

$$\frac{\partial n}{\partial t} = D\left(\frac{\partial^2 n}{\partial x^2} + \frac{\partial^2 n}{\partial y^2}\right). \tag{1}$$

The equation expresses the fact that the rate of change of the cell density over time ($\partial n/\partial t$) is related to the diffusivity $D$ and the variation of $n$ in space (the coordinates $x$ and $y$). Further details of how the equation is derived from random walk theory and its extensions are presented in Ref. [45]. It is always useful to consider the differential analogue of agent-based probabilistic models (and vice versa) as different applications may favour different mathematical approaches. The diffusion equation and isotropic random walks form the basis of more complex migratory models, including biased and correlated random walks discussed below.

### 2.1.2 Biased and correlated random walks

The concept of an isotropic random walk (also known as Brownian motion) provides a flexible basis for a detailed characterisations of cell movements. Remarkably, it can be naturally extended to a wide range of circumstances. For example, cells often have a preferred direction, e.g., in response to differences in their environment. Random walks can be biased by an external source giving preference to movement in a particular direction (a biased random walk or BRW) [37]. This bias can be accounted for by making the step length in the preferred direction longer or by increasing the probability that the preferred direction is chosen. The BRW trajectory shown in Figure 4 is biased towards the right (in this case the step length remains the same in every direction but it is more likely that a movement to the right will occur than any other). BRWs can be a suitable model for cells in the presence of biasing chemicals or boundaries [24, 46-48].

As the IRW leads to the diffusion equation, the BRW also has an analogous differential equivalent: the drift–diffusion (or advection–diffusion) equation

$$\frac{\partial n}{\partial t} = D\left(\frac{\partial^2 n}{\partial x^2} + \frac{\partial^2 n}{dy^2}\right) - v_x\frac{\partial n}{\partial x} - v_y\frac{\partial n}{\partial y}. \quad (2)$$

The first term in the equation represents the diffusive part of the movement as in Eq. (1) with the second and third terms capturing the drift (bias) in the $x$ and $y$ directions at the speeds $v_x$ and $v_y$, respectively. Such diffusion equations are not only used in migration models, but also for descriptions of chemotaxis [49] and ecological models [50].

Another generalisation of the IRW is a correlated random walk (CRW) where the direction of each step depends on the previous step directions. The walk can be persistent, where the next step is more likely to be close to the direction of the previous step, or anti-persistent, where the next step is more likely to be in the opposite direction. The example CRW trajectory shown in Figure 4 is persistent, the cell prefers to keep going in the direction in which it has just been, and so the trajectory ends up more directed than for isotropic movement. CRWs often occur in cell kinematics in the absence of external biases [51-53]. More generally, animal movements often take the form of a CRW (e.g., cabbage butterflies [54], caribou [55] and seals [56]).

*Figure 4 – Trajectories of isotropic (blue), biased (orange) and correlated (yellow) random walks. All trajectories start from the same position (the black circle). The biased trajectory is extended towards the right and the correlated trajectory has relatively long parts with large directionality.*

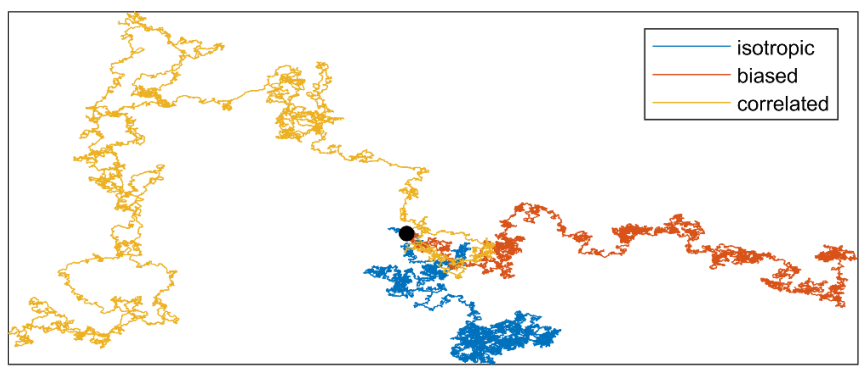

### 2.1.3 A note on circular statistics

When considering the directional migration of cells, the quantities of interest often include statistical properties (e.g., averages and standard deviations) of angular measures. The averaging of circular quantities (circular statistics), such as angles, has some peculiar properties. For example, consider taking the mean of the angles $\theta_1 = 0°$ and $\theta_2 = 270°$, shown in Figure 5. By intuitively visualising the average direction of the two angles, the answer is $\bar{\theta} = 315°$. However, the arithmetic mean is $\bar{\theta} = (\theta_1 + \theta_2)/2 = 135°$, not the intuitive value or representative directional average in this case as it differs by $180°$ from the correct value $\bar{\theta} = 315°$.

*Figure 5 – Consider taking the average of two angles, $\theta_1 = 0°$ and $\theta_2 = 270°$ (blue). Using the arithmetic mean (left) gives $\bar{\theta} = 135°$ (orange) while using the circular mean (right) gives the more intuitive result $\bar{\theta} = 315°$.*

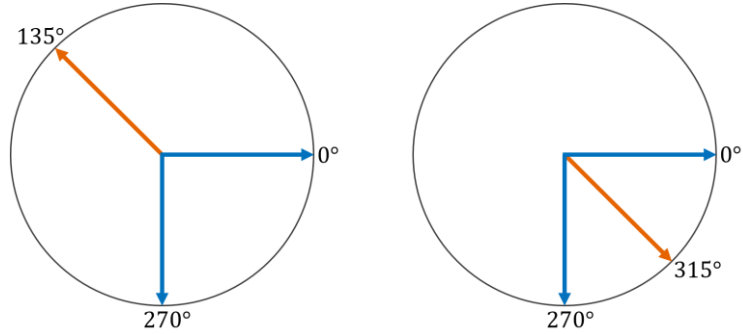

To get the correct value one must use the circular mean, defined for angles $\alpha_1, \ldots, \alpha_n$, as

$$\bar{\alpha} = \text{atan2}\left(\sum_{j=1}^{n} \sin(\alpha_j), \sum_{j=1}^{n} \cos(\alpha_j)\right),$$

where atan2 is a special trigonometric function, similar to the usual arctangent, except the signs of both arguments are used to determine the quadrant of the result,

$$\text{atan2}(y, x) = \begin{cases} \arctan\left(\frac{y}{x}\right) & \text{if } x > 0 \\ \frac{\pi}{2} - \arctan\left(\frac{x}{y}\right) & \text{if } y > 0 \\ -\frac{\pi}{2} - \arctan\left(\frac{x}{y}\right) & \text{if } y < 0 \\ \arctan\left(\frac{y}{x}\right) \pm \pi & \text{if } x < 0 \\ \text{undefined} & \text{if } x = 0 \text{ and } y = 0. \end{cases}$$

In other words, the quantities amenable to averaging are not the angles but their tangents (calculated with due allowance to the quadrants where the angles lie, e.g., the standard function atan2 rather than atan in Matlab).

Similarly, the circular correlation between two angular quantities is defined as

$$C = \frac{\sum_i \sin(\alpha_i - \bar{\alpha})\sin(\beta_i - \bar{\beta})}{\sqrt{\sum_i \sin^2(\alpha_i - \bar{\alpha})\sin^2(\beta_i - \bar{\beta})}},$$

where $\alpha_i$ and $\beta_i$ denote two samples of angles and $\bar{\alpha}$ and $\bar{\beta}$ their angular means. Further details of the use of circular statistics for biological applications can be found in Ref. [57]. Matlab users may also find the circular statistics toolbox (directional statistics) useful [58]. There are similar resources available for R [59].

## 2.2 Stem cell kinematics

Having introduced the concept and classification of random walks, we now consider their application to stem cell migration. A thorough understanding of the migration of hPSCs is needed to optimise *in-vitro* clonality and facilitate the development of therapies for migration related disorders. Unregulated cell migration *in-vitro* can cause clonality loss as cells of different lineages aggregate into groups which is undesirable when a genetically identical clonal population is required [60, 61]. Furthermore, anomalous cell migration has been linked to deviations in the undifferentiated state of iPSCs [62]. Here we discuss the kinematics of isolated cells and their pairs *in-vitro*.

Individual cell movements have been explored through direct experiments with hPSCs (in particular human embryonic stem cells, hESCs) and analysed within the random walk framework described in Section 2.1 [16, 24, 25]. The movement of single hESCs can be accurately described as an isotropic random walk when the cells are in isolation, i.e., more than approximately 150μm away from any neighbouring cells. As the separation distance decreases, the cell movements become more directed towards each other, with motility-induced aggregation occurring in 70% of instances when the distance been two hESCs is less than 6.4μm [24]. A minority of isolated single cells exhibit super-diffusive behaviour (and so explore their environment faster), contributing heavily to the motility related clonality loss [16, 24, 30]. Example experimental trajectories for cells exhibiting typical diffusive and super-diffusive migrations are shown in Figure 6. These results show that individual cell movement influences hPSC clonality, although particular biological drivers of the distinct diffusive behaviours remains unclear. Further research in this direction can provide guidance for improvement of clonogenic assays in the analysis of hPSC self-renewal [24] and be used to identify timescales for motility-driven cross-contamination between colonies which is of practical use when producing high clonality colonies.

*Figure 6 – Example cell trajectories for isotropic motion in-vitro around a central point (left) and a directed walk (right). The initial and final cell centroid positions are shown as a circle and a square respectively (note that these points are not representative of cell or nucleus size).*

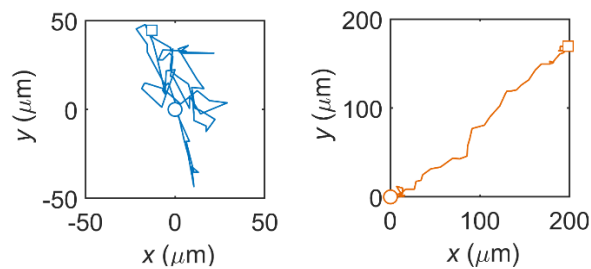

Experimental analysis of hESCs [25] has shown evidence of correlated random walks of individual isolated stem cells. Single hESCs (more than 150μm away from any neighbouring cells, as in [24]) tend to perform a locally anisotropic walk, moving backwards and forwards along a preferred local direction correlated over a time scale of around 50 minutes, becoming more persistent over time. The preferred direction of motion is aligned with the axis of cell elongation (Figure 7) which could reflect an optimised strategy of search for neighbouring cells. Further experiments are required to understand how the presence of multiple neighbours affects this anisotropic movement.

*Figure 7 – A single hESC migrating backwards and forwards along a local axis (left). The blue dot shows the cell nucleus and the black arrow the direction of instantaneous velocity. The scale bars are 30μm in length [25]. A polar histogram of the direction of motion (relative to the last step direction) of experimental hESCs (right) [25] with the red line indicating the average preferred direction of movement. The cells show a preference for moving either in the same direction, or the reverse direction to the previous step.*

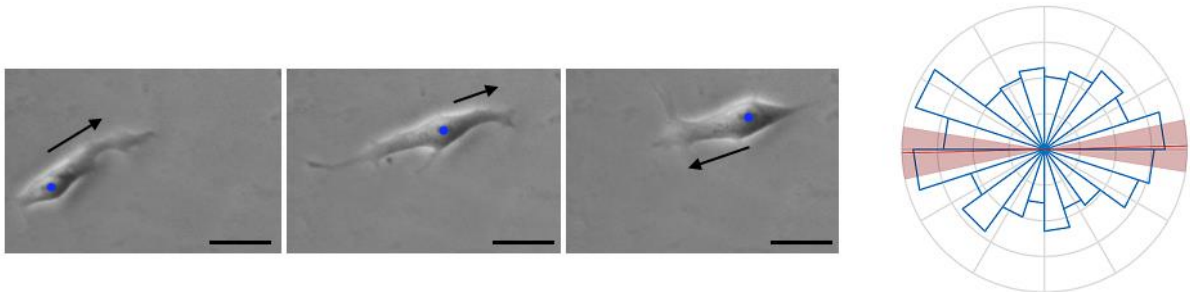

Pairs of hESCs tend to move in the same direction, with the average separation of 70μm or less and a correlation length (the length scale of communication) of around 25μm. Pairs often remain connected by their pseudopodia, even at larger separations (>100μm) when they exhibit independent movements. For the correlated pairs, it is not known whether the movement correlation is facilitated by the physical connection or the coordination is due to inter-cell chemical contacts alone.

When cell motility has been quantified, the direct comparison of cell behaviour for distinct cell types and under different experimental conditions becomes possible. For example, the addition of a CellTracer™ (a marker for the tracking of cell generations) results in significantly reduced migratory behaviour for individual and pairs of cells [16], and thus, can affect their potential to develop into colonies as well as the colony clonality.

There is evidence that cell migration in 3D cannot be described as a persistent random walk and new models need to be developed to accurately describe this motion [40]. These experimental results further inform the development of individual-based models that can be integrated into more complex models of cell movement within colonies (as discussed in Section 4.1).

# 3 Differential equations

*Synopsis: The aim of this section is to introduce readers to the concept of differential equations which are used to describe the rates of change of physical properties. We discuss the exponential and logistic growth models, commonly used to describe population numbers, and how these models are being used to improve human pluripotent stem cell clonality. We also present some recent examples of differential equations for describing cell pluripotency.*

Differential equations (DEs) are a fundamental part of the mathematical descriptions for all biological systems. They describe the relationship between the properties of the cells, their environment, and their rates of change in space and time. In Section 2.1 we encountered two DEs from the theory of random walk models: the diffusion equation and the drift-diffusion equation. This section introduces some further common DEs used in mathematical biology, discusses the differences between deterministic and stochastic DEs, and presents some pertinent examples relevant to stem cell biology.

## 3.1 Fundamental equations

DEs describe rates of change, for example, the rate of temporal change is the time ($t$) derivative. For example, the rate of change of the number of cells ($N$), written as $dN/dt$ (or $\partial N/\partial t$ if $N$ depends on more variables than just $t$), is given by the ratio of increment in $N$, $\Delta N$, to the time interval $\Delta t$ over which it has been accumulated, $dN/dt \approx \Delta N/\Delta t$. This equality is only approximate because $dN/dt$ is the instantaneous rate of change in $N$, whereas $\Delta N/\Delta t$ is the rate averaged over the time interval $\Delta t$. The shorter is $\Delta t$, the close is $\Delta N/\Delta t$ to $dN/dt$. To illustrate the use of this quantity, consider a population of $N$ cells, with a fraction $r$ of cells dividing in unit time. Then $\Delta N$ is proportional to both $r$ and $N$ and increases with the time $\Delta t$ of the evolution: $\Delta N = rN\Delta t$. Thus $\partial N/\partial t \approx \Delta N/\Delta t$ and the DE describing the evolution of the population is $dN/dt = rN$, meaning that the instantaneous differential (increment) of $N$ per unit time is given by $rN$. This equation can be further enriched by including various effects that influence the cell division, as we discuss below. This equation has the solution of the form $N = N_0 e^{rt}$ where $N_0$ is the initial size of the population (at $t = 0$, the starting time of the evolution) and $e \approx 2.7$, Euler's number. The evolution of $N$ over time for different initial population sizes and different growth rates are shown in Figure 8. Negative values of $r$ describe a dying population and positive value correspond to its proliferation. Below we discuss an extension of this, the logistic equation.

*Figure 8 – Left: The time evolution of a population N following exponential growth, $N_0 e^{rt}$, with $r = 0.1$ cells/minute and $N_0 = 1$ (blue solid), 2 (orange dashed) and 3 (yellow dotted). Right: Exponential growth with $N_0 = 1$ and $r = 0.05$ (blue solid), 0.1 (orange dashed) and 0.15 cells/minute (yellow dotted).*

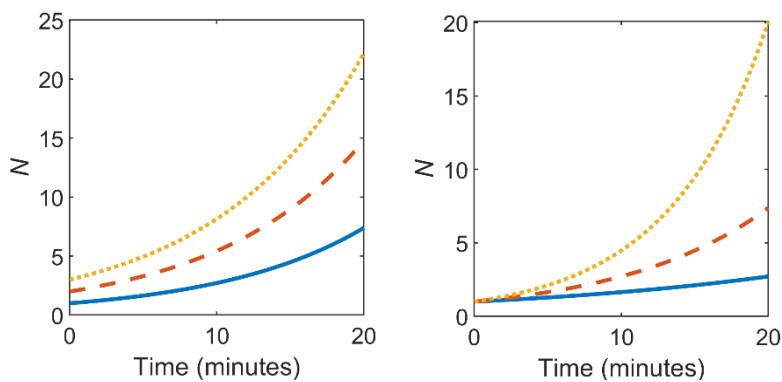

There are two major types of differential equation: determinist and stochastic. Deterministic DEs have no randomness in their parameters (such as the growth rate $r$) and so have the same solution for a given initial state. Stochastic DEs (or SDEs) include randomness in the parameters, which results in different solutions even for the same initial state. Real-life biological scenarios often include some element of randomness, and it is for this reason that SDEs are often chosen as better models. We will use the deterministic and stochastic logistic equation, used to model population growth, to illustrate the differences between deterministic DEs and SDEs.

The logistic growth model is fundamental in cell biology and ecology. The equation describes the evolution of population size $N$ with time $t$ as

$$\frac{dN}{dt} = rN\left(1 - \frac{N}{K}\right), \tag{3}$$

where the growth rate of the population is $r$ and $K$ is the carrying capacity, which represents the maximum population size allowed in a given environment. As $N$ increases, the instantaneous growth rate $r(1 - N/K)$ decreases until the growth stops when $N = K$ and the population settles into a steady state. Examples of a population growing according to Eq. (3) are shown in Figure 9. This equation is deterministic, solving the equation again with the same initial condition $N_0$ gives the same results.

The stochastic logistic equation adds randomness to the standard logistic model of Eq. (3). Some element of randomness, or noise, can be essential in many circumstances. The population then broadly follows the logistic equation, Eq. (3), but only on average. This approach is based on the concept of averaging over different realisations of the population growth that have slightly different environments and individual cell properties. Each realisation can represent a separate experiment. It is impossible to reproduce the system with perfect accuracy in different experiments. This uncertainty leads to the randomness. One of the fruitful stochastic logistic models is where the overall rate of change of $dN/dt$ fluctuates:

$$\frac{dN}{dt} = rN\left(1 - \frac{N}{K}\right) + \sigma W_t, \tag{4}$$

where the additional noise term is given by $\sigma W_t$ (this is known as the Wiener process, purely random noise analogous to the Brownian motion) with the scaling parameter $\sigma$ which determines the strength of the noise and $W_t$ which captures the statistical properties of the noise (e.g., Gaussian). Examples of the evolution described by Eq. (4) are shown in Figure 9. Although each of the three solutions follow the same trend, i.e., the average behaviour, the noise in the system results in three different evolution trajectories. The scaling $\sigma$ controls the strength of the random variation; the greater $\sigma$ the greater the variation in the solutions.

*Figure 9 –The population size growing with time as described by the deterministic logistic equation (3) (left) with the growth rate $r = 0.1$ cells/minute, carrying capacity $K = 10$ cells, and the initial population sizes of one (blue, solid), two (orange, dashed) and five (yellow, dotted) cells. Different realisations of the solutions for the stochastic logistic equation (4) (middle) with the noise magnitude $\sigma = 0.1$. The blue solid line in the left panel represents the mean value of such realisations. Solutions to the stochastic logistic equation with increasing noise (right); $\sigma = 0.1$ (blue, solid), 0.2 (orange, dashed) and 0.5 (yellow, dotted).*

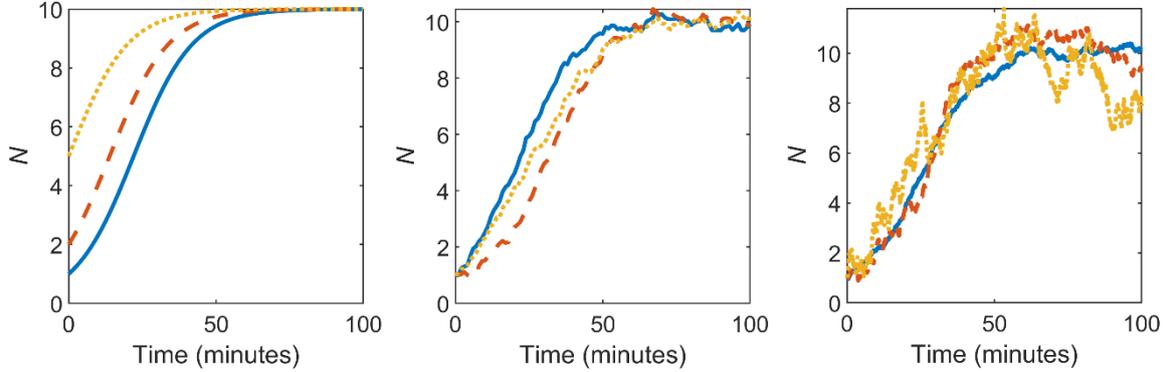

The randomness of real systems can affect not only the overall rate of growth $dN/dt$, as in Eq. (4) but also the intrinsic growth rate $r$. This source of randomness reflects effects such as variation in the length of the cell cycle from cell to cell. Thus, the intrinsic growth rates now contain a random part, becoming $r + \sigma W_t$. There are other adaptions which include a time-dependent carrying capacity, the Allee effect where the population growth rates are reduced at low population size and time delays where changes in the system parameters affect the population size only after a certain delay. Stochastic logistic equations have been used widely, particularly in ecology and cell biology. The next sections review some examples of differential equation models which have been developed to capture stem cell behaviour, particularly pluripotency and colony growth due to cell proliferation.

### 3.2 Colony growth

Population models have been used to understand the process by which blood cells are formed [63], cancer tumours grow [64] and the impact of hPSC colony growth on clonality [7]. Early population dynamics models for stem cells were based on stochastic birth-death processes [18] involving systems of ordinary differential equations [65] similar to the logistic equation discussed above. One of the most popular models for hPSCs allows for dividing and non-dividing cells, and incorporates cell loss through death or differentiation (often referred to as the Deasy model, which is a development of the Sherley model to include cell loss) [66, 67]. The evolving number of cells over time $N(t)$ is obtained as

$$N(t) = N_0 \left[ \frac{1}{2} + \frac{1 - (2\alpha)^{t/T+1}}{2(1 - 2\alpha)} \right] - M, \quad (5)$$

where $N_0$ is the initial number of cells, $\alpha$ is the mitotic fraction (the fraction of the cells that can divide), $T$ is the cell cycle duration, and $M$ is the number of cells lost to death or differentiation. Since the model accounts for both differentiating and pluripotent cells, it can be used to assess the heterogeneity of a cell population and the evolution of mixed subpopulations [66]. The model has been compared with experimental results for muscle-derived stem cells [66] and hESCs [68].

More recently, a new family of growth models has been developed, known as hyperbolastic growth models. These models are designed to capture self-limiting growth behaviours [69] and have been introduced for both adult and embryonic stem cells [68]. The hyperbolastic growth models provide more flexibility in the growth rate as the population reaches its carrying capacity and have been demonstrated to capture experimental data well [68, 69]. The population $N$ in this case is governed by the differential equation

$$\frac{dN}{dt} = (K - N)\left[r\gamma t^{\gamma-1} + \frac{\theta}{\sqrt{1 + \theta^2 t^2}}\right], \qquad (6)$$

describing the rate of change in the number of cells, $dN/dt$, starting from the initial condition $N = N_0$ at $t = 0$. The parameters represent the salient properties of the system: $K$ is the carrying capacity of the environment, $r$ is the intrinsic growth rate, the factor $t^{\gamma-1}$ with a certain constant $\gamma$ is responsible for a decrease ($\gamma < 1$) or increase ($\gamma > 1$) in the cell proliferation rate with time, and $\theta$ allows for further variation in the instantaneous growth rate. This model can be used to describe both proliferation and cell death rates more accurately than Eq. (5) [68] and helps identify when the growth of cells becomes self-limiting, a biological problem currently not fully understood.

Experimental data is essential to inform model development. For hPSCs, the parameters of an exponential growth model (as discussed in the introduction to this section) have been elucidated from experimental data [7]. Colony populations are multi-modal, suggesting a group of colonies formed from a single cell (with smaller colony sizes) and a group formed from pairs of cells (with larger colony sizes), importantly showing inherent differences in the biological behaviours of cells with different numbers of neighbours [7]. The corresponding model is based on a deterministic exponential growth model with growth rate $r$, $dN/dt = rN$, and solution $N = N_0 e^{rt}$ for $N_0$ initial cells. Since the cell cycle duration is variable due to various factors (e.g. inhomogeneities in the nutrient distribution within the growth medium and the inherent variation in the cell cycle between different clones), it is appropriate to make the equation stochastic. Randomness can be incorporated into the growth rate $r$ by sampling $r$ randomly from a Gaussian distribution: $r \sim \text{Norm}(\mu, \sigma^2)$. The parameters of this distribution are found from the experimental data. Since there are two populations ($N_A$ and $N_B$) with two different initial conditions (one founder cell colonies with probability $\alpha$ and pairs of cell founder colonies with probability $\beta$), two equations are required:

$$\begin{cases} N_A = e^{r_A t}, \ r_A \sim \text{Norm}(\mu_A, \sigma_A^2), \text{ with probablity } \alpha \\ N_B = 2e^{r_B t}, \ r_B \sim \text{Norm}(\mu_B, \sigma_B^2), \text{ with probablity } \beta \end{cases} \qquad (7)$$

with $\mu_A = 0.039$ and $\sigma_A^2 = 0.006^2$, $\mu_B = 0.043$, $\sigma_B^2 = 0.002^2$, $\alpha = 0.77$ and $\beta = 0.23$ inferred from the fitting to the experimental data. The growth rate for colonies emerging from pairs of cells is $\mu_B$ greater on average than for colonies founded by single cells, $\mu_A$. This means that colonies that have grown from cell pairs are larger not only due to the initial condition but also because their proliferation rate is often larger. Moreover, the scatter in the mean growth rates is significantly smaller in the case of founding pairs, $\sigma_A/\sigma_B = 3$. These features are consistent with observations that hPSCs proliferate more effectively when in close proximity to other cells [70, 71]. This difference is important when the clonality of a colony needs to be assessed non-invasively, e.g., from its size. Finally, the values of $\alpha$ and $\beta$ reflect the probabilities for a cell to occur in isolation or in a pair when seeded at low average density.

Upon collection of further experimental data, the model can be expanded to describe colony growth from larger groups of founder cells. It is expected that the growth rate for colonies will increase with the number of starting cells before reaching its peak and needs to be quantified. These growth rates may vary under different experimental conditions.

The model can be used to predict hPSC colony growth and to estimate the time lag over which the colony size no longer reflects the number of founding cells. Up to this critical time, the colony size can be used as a non-invasive marker of clonal colonies that have grown from a single cell.

The multi-population model of Eq. (7) can be expanded to simulate colony growth in space and used to explore the impact of colony growth on the clonality [7]. Generating homogeneous populations of clonal cells is of great importance [60, 61] as clonally derived stem cell lines maintain pluripotency and proliferative potential for prolonged periods [72]. To achieve this, the cross-contamination and merging of growing colonies (illustrated in Figure 10) should be avoided.

*Figure 10 – Left: An example of two colonies merging from experimental images. The two colonies, shown in blue and green are beginning to merge five days after seeding. The scale bar represents 100μm. Right: Diagram illustrating initially seeded cells (filled circles) growing into colonies (dashed) and the first time at which the two growing colonies touch each other from a simulation of the cell seeding model of Eq.(7). The cells shown in green represent a pair and proliferate accordingly faster producing a larger colony (shown by a green circle in the far right panel). From Ref. [7].*

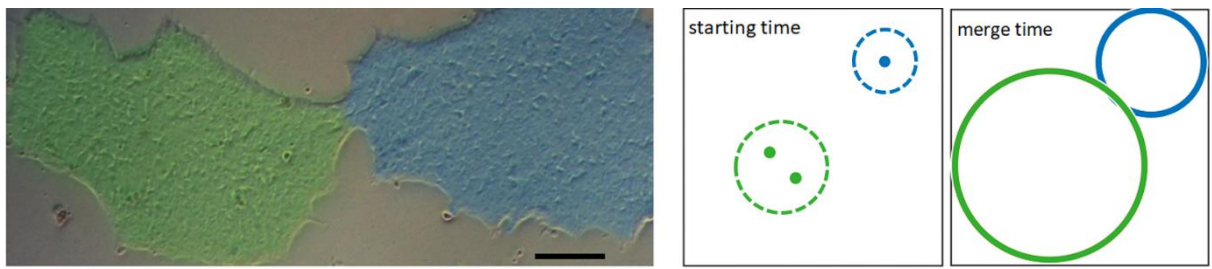

Assuming that, initially, the cells are randomly scattered in a growth area with a particular seeding density (the average number of cells per unit area), each cell (or group of cells) proliferates according to Eq. (7). Each colony is then approximated by a circle, with a certain position in space (the geometric centre of the locations of the founding cells) and a radius based on the population size and an assumed cell area of $250\mu m^2$ per cell [33]. The time at which a colony begins to merge with its neighbour, $T$, is the time at which the perfect clonality is lost as illustrated in Figure 10. Simulating colony growth for different seeding densities leads to an equation for the critical time $T = (-0.007 \pm 0.0001)n_0 + (102 \pm 3)$ with $T$ in hours and $n_0$ the initial seeding density of cells before their attachment to the substrate in cells/cm$^2$. Therefore, the larger is the seeding density $n_0$, the sooner the colonies merge leading to the loss of their clonality. For example, the seeding at the average density $n_0 = 10^4$ cells/cm$^2$ would lead to the loss of clonality after $T \approx 102 - (10^4/140) \approx 30$ hours of evolution. This application of mathematics has direct implications for experimental optimisation: the key equation can be used to achieve the best outcome for homogeneous colony growth *in-vitro* by choosing the optimal cell seeding density.

### 3.3 Pluripotency modelling

Mathematical models of pluripotency are deepening our understanding of how pluripotency is regulated, leading to the optimisation and control of cell pluripotency in the laboratory. Modelling pluripotency and cell fate decisions remains challenging as even clonal cells under the same conditions make different fate decisions, and it remains unclear how much fate choice is led by inherited factors versus environmental factors and intracellular signalling [73]. There are several thorough reviews of the computational models of cell fate decisions [74-76]. Here we focus on a few current models that aim to capture the fluctuating pluripotency in individual cells over time.

At the single-cell level, pluripotency is inherently stochastic; indeed, it has been proposed that pluripotency is only defined statistically within a population [13]. Colonies exhibit heterogeneous subpopulations of cells with differing levels of pluripotency transcription factor (PTF) expression [14, 77]. This suggests that the evolution of cell colonies involves a competition between disruptive single-cell influences and regulatory community effects. There is recent evidence that differentiation potential is linked to the length of the G1 phase of the cell cycle [78]. Such heterogeneity is undesirable as it can bias evolution trajectories and lead to spatially disordered differentiation [79, 80]. Here we will consider intra-cellular models of pluripotency that focus on the inter-connected dynamics of OCT4, Nanog and SOX2, the main PTFs.

A widely used model of the intra-cellular PTF network regulation involves the Hill equation [81, 82] of the theory of chemical reactions. In the simplest form, the concentration of a reacting substance $X$ activated by the transcription factor $Y$ is described as

$$\frac{dX}{dt} = r \frac{Y^n}{K^n + Y^n},$$

where $r$ is the maximum transcription rate, $K$ is the value of $Y$ at which the rate of change of $X$ is reduced to $r/2$ (the so-called equilibrium constant) and $n$ (the Hill coefficient) characterizes the sensitivity of the dynamics of $X$ to the magnitude of $Y$: the larger is $n$, the higher is the sensitivity. This equation can be generalised and extended to describe a complex network of mutually regulating chemical reactions. The model of Ref. [81] describes the rate of change of each transcription factor in the network with the equation for the concentration of OCT4 ($O$) in relation to Nanog ($N$), SOX2 ($S$) and OCT4-SOX2 ($\theta$) given by

$$\frac{dO}{dt} = \frac{\eta_1 + a_1 A_+ + a_2 \theta S + a_3 \theta N}{1 + \eta_2 + b_1 A_+ + b_2 \theta + b_3 \theta N} - \gamma O - k_1 OS + k_2 \theta, \qquad (8)$$

where $\eta_1$ and $\eta_2$ are the basal transcription rates, $A_+$ is a signal positively regulating the levels of OCT4, $\gamma$ represents the decay rate, $k_1$ and $k_2$ are the kinetic constants and $a_{1,2,3}$ and $b_{1,2,3}$ are the constants related to the free binding energies. The large number of parameters that appear in the resulting system of four equations, which are difficult to estimate accurately, diminishes the advantages of this approach [74, 83].

The Nanog regulation of OCT4 and SOX2 has also been considered using similar differential equations [84]. This model considers two different regimes for Nanog expression: stochastic fluctuations and deterministic oscillations. For stochastic fluctuations alone, a Gaussian white noise is included in the differential equation for Nanog concentration ($N$) in relation to OCT4-SOX2 ($\theta$)

$$\frac{dN}{dt} = \frac{s_4 N^{\gamma_N}}{k_4 + N^{\gamma_N}} + \frac{s_3 \theta^{\gamma_O}}{k_3 + \theta^{\gamma_O}} - d_N N + N \zeta(0, \sigma), \qquad (9)$$

where the coefficients $s_3$ and $s_4$ denote the transcription rates from the OCT4-SOX2 complex and self-influence, respectively, $K_3$ and $K_4$ are the equilibrium constants, $\gamma_N$ and $\gamma_O$ are the Hill coefficients and $\zeta$ denotes the random contribution proportional to the concentration of Nanog $N$ whose magnitude is controlled by the standard deviation of the Gaussian noise $\sigma$. The term with $d_N$ describes the natural degradation of Nanog. This regime results in the dynamics shown in Figure 11. For certain parameter values, the Nanog concentration tends to stabilise at either a low or high level but the random perturbations induce transitions between the two states. In the other regime, which reproduces oscillatory behaviours, the governing equation includes a competitive repression of Nanog due to a transcriptional repressor $X$,

$$\frac{dN}{dt} = \frac{s_4 N^{\gamma_N}}{k_4 + N^{\gamma_N} + s_6 X^{\gamma_X}} + \frac{s_3 \theta^{\gamma_O}}{k_3 + \theta^{\gamma_O}} - d_N N + N\zeta(0, \sigma), \qquad (10)$$

while the repressor dynamics are governed by a similar equation,

$$\frac{dX}{dt} = \frac{s_5 N^{\gamma_N}}{k_5 + N^{\gamma_N}} - d_X X. \qquad (11)$$

The dynamics for Nanog in this regime are also shown in Figure 11. The oscillatory system is characterised by a limit cycle where the system converges to a predictable, periodic state, independent of initial state. Although indistinguishable on the cell population level, the two mathematical models suggest quite different inherent Nanog variation, and further experimental testing is required to assess the applicability of these models. There are many other approaches to the PTF modelling using differential equations [85-90].

Models of this kind can reproduce a wide range of complex behaviours and potentially include a large number of various factors that affect the bio-chemical networks. However, the usefulness of this approach is strongly limited by the fact that the numerous parameters that appear in the resulting equations are difficult or impossible to determine with any degree of accuracy and generality. On the other hand, the very fact that the number of factors that affect the PTF dynamics is so large, opens up the possibility to describe the bio-chemical regulation as a random system where the complex network of interacting positive and negative feedback loops leads to dynamics captured by a few universal parameters. The situation is not dissimilar to the description of a complex random motion in terms just a single parameter, the diffusivity $D$ as described in Section 2. In other words, beyond a certain level of complexity, the dynamics of bio-chemical networks is amenable to the stochastic chemistry approaches.

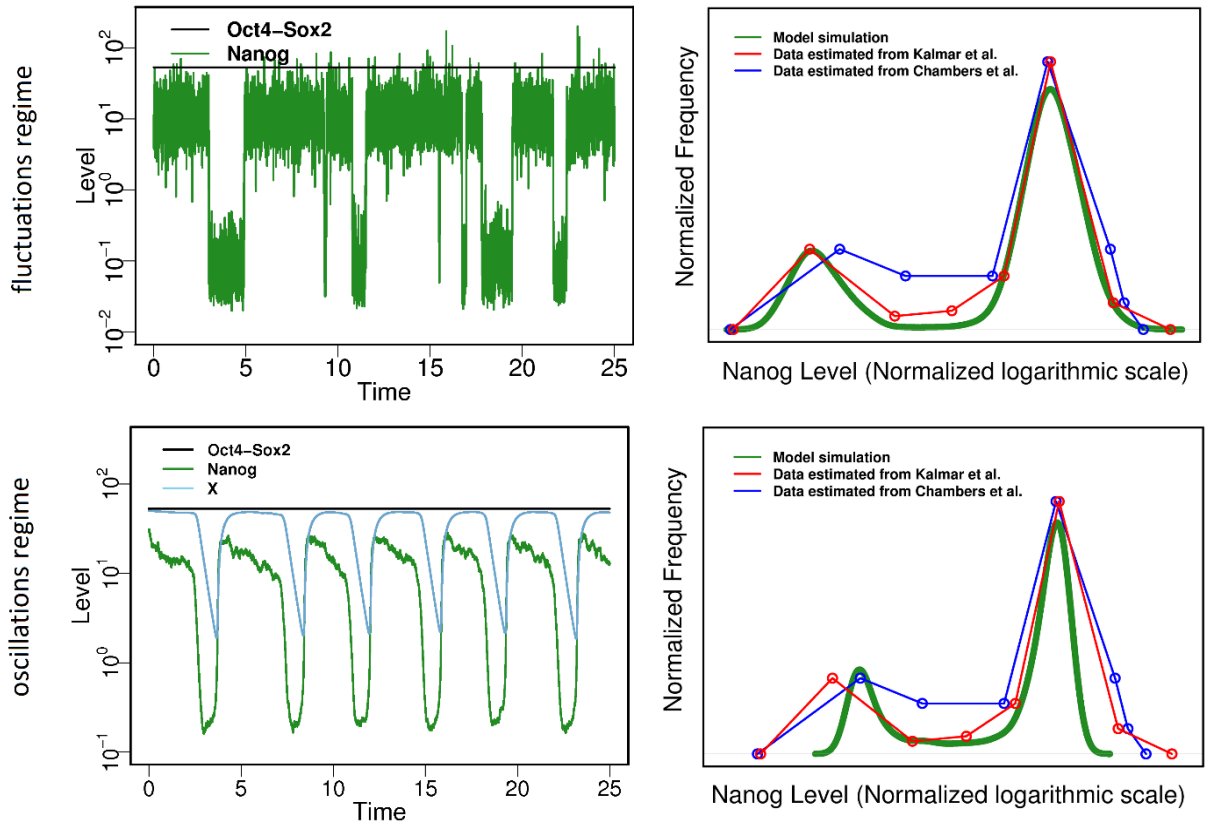

*Figure 11 – Left column: The time variations of the concentrations of the PTFs Nanog (green) and OCT4-SOX2 (black) in the fluctuation (top row) and the oscillation (bottom row, with the dynamics of X shown with blue line) regimes as obtained from Equations (9)-(11). Right column: the comparison of the occurrence frequency of the Nanog concentrations in these models and the experiments of Ref. [85] and Ref. [91]. Figure adapted from Ref. [84].*

Pluripotency also shows characteristic spatial variation on the colony scale. Preliminary experiments monitoring the levels of OCT4 in single hESC founder cell colonies at 72h post seeding show that pluripotency is clustered, with highly pluripotent cells grouped together, shown in Figure 12.

The differentiation of hPSCs also shows distinctive and apparently systematic spatial patterning [79, 80]. The spatial distributions of the pluripotency marker SOX2 and the differentiation marker AP2α suggest that the differentiation occurs preferentially at the colony periphery in a band of a constant width, about $150\mu$m independent of colony size as shown in Figure 12 [66] and illustrated schematically in the rightmost panel of Figure 1. The differentiated cells originate in the outer third of the colony and remain near its boundary as the colony grows. This provides important information for modelling the spatial patterning of the pluripotent state.

*Figure 12 – Left: A microscopy image of a hESC colony at 72h after seeding, alongside a colour-coded version of the same colony quantifying the level of expression of OCT4. Red represents the highest pluripotency with blue representing the lowest. Scale bar represents 50μm. Right: Phase (top) and immunostaining images (bottom) of hESC colonies before and after BMP4 addition showing a band of differentiation. Figure adapted from Ref. [80].*

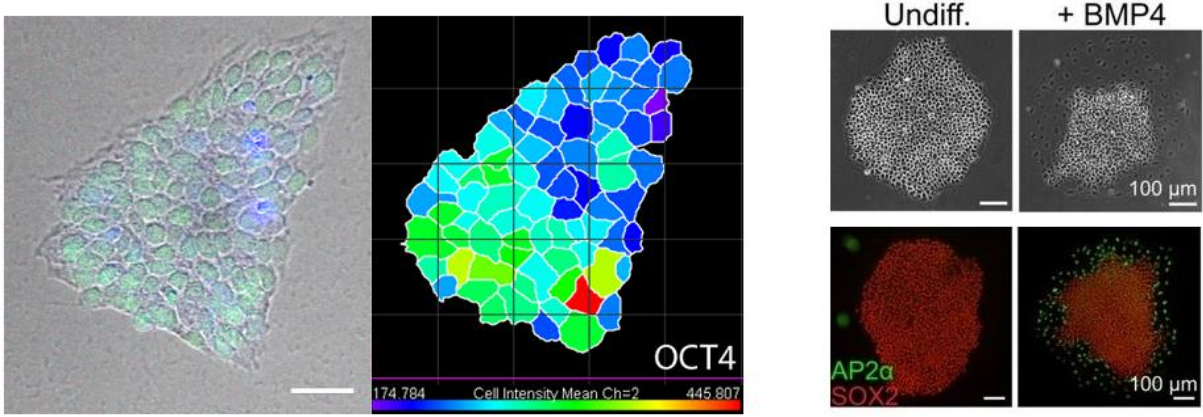

The spatial segregation of differentiated cells is captured in a model of cell movements in a colony which is based on the balance of the intracellular and extracellular mechanical forces [92]. This is known as a mechanical bidomain model and was first developed to describe the elastic behaviour of cardiac tissue [93]. In agreement with the experimental results [79, 80], the model predicts that differentiation and traction forces occur within a few length constants of the colonies edge. The length scale is determined by the intracellular and extracellular cell rigidities (shear moduli) and the coupling of the internal and external mechanical forces. The model assumes that differences in displacement between the pluripotent and differentiated cells are responsible for the mechano-transduction (chemical processes through which cells sense and respond to mechanical stimuli). Equations for the intracellular ($u_r$) and extracellular ($w_r$) displacements, functions of the distance from the colony centre ($r$), have the form

$$4\nu \left( \frac{\partial^2 u_r}{\partial r^2} + \frac{1}{r}\frac{\partial u_r}{\partial r} - \frac{u_r}{r^2} \right) = K(u_r - w_r),$$

$$4\mu \left( \frac{\partial^2 w_r}{\partial r^2} + \frac{1}{r}\frac{\partial w_r}{\partial r} - \frac{w_r}{r^2} \right) = -K(u_r - w_r),$$

(12)

where $\nu$ and $\mu$ are the intracellular and extracellular shear moduli, respectively, and $K$ is their coupling constant. The difference between the intra and extra-cellular displacements follows as

$$u_r - w_r = -\frac{T\sigma}{4\nu} e^{\frac{r-R}{\sigma}},$$

(13)

where $T$ is a uniform stress caused by the growth and crowding of cells, $\sigma = \sqrt{4\nu\mu/[K(\nu+\mu)]}$ is the characteristic length and $R$ is the colony radius. As shown by this relation, the difference is very sensitive to the distance $r - R$ to the colony boundary and rapidly decreases away from the boundary at the length scale $\sigma$ which is independent of the colony radius $R$.

This model assumes that the difference between the intra-cellular and extra-cellular forces is the primary driver (or evidence of) the differentiation. This model could be further developed to allow for an irregular colony shape. It is worth investigating whether the cell growth, represented by the tension $T$, is a function of $u_r - w_r$ alone (as assumed in the model), since hESC observations suggest distinct actin organisation and greater myosin activity near the colony edge, implying that $T$ could be non-uniform [80].

Further experiments are needed to explore the pluripotency in various types of colony. In addition to those already mentioned, thorough reviews of other mathematical models of pluripotency are available [19, 94], along with a review of computational modelling of the fate control of mouse embryonic stem cells, with many models transferable to iPSCs [74]. The mathematical exploration of pluripotency promises a better understanding of pluripotency regulation and differentiation within colonies.

# 4 Agent-based modelling of colonies

*Synopsis: The aim of this section is to introduce the concept of agent-based modelling for cellular systems. We discuss how agent-based models are used to describe cell migration and colony proliferation.*

The understanding of larger groups and colonies of hPSCs is complicated by the fact that they involve both collective and individual behavioural effects [95]. For example, the coordinated migration of large numbers of hPSCs *in-vivo* is essential in tissue generation [96] and wound healing [97], and as discussed in Section 3.3, pluripotency shows spatial patterning on the colony scale [79, 80]. A flexible approach to modelling large colonies of cells with allowance for their complicated multi-scale behaviours is the agent-based modelling technique.

Agent-based models (ABMs, sometimes referred to as individual-based models or IBMs) consist of individual components, called 'agents', which are characterized by their locations and behave according to a set of rules that reflect their response to the environment and internal processes. This type of a model is based not on differential equations where the equation parameters reflect our knowledge of the cell properties, but rather on a set of explicit rules for the evolution of each 'agent' in response to various stimuli. It is often convenient to think of an 'agent' as a single cell although it can represent a group of cells if appropriate. The governing rules are often assumed to be probabilistic. The rules of behaviour usually include the probability of movement in any of the allowed directions (the probabilities often depend on the state of the neighbouring cells) and changes in the 'agent' state depending on time or the neighbouring 'agents' (implementing, for example, the cell cycle). ABMs are commonly used in the studies of collective behaviours in mathematical biology and population dynamics not only for cells but also for schools of fish [98] and crowds of people [28, 99].

Each ABM 'agent' can represent an individual cell in a colony. The set of rules that describe the behaviour of the cells and the colony as a whole can include a wide variety of external and internal factors leading to complex and sophisticated models. For example, each cell's individual migration could occur as an isotropic random walk (in fact, the random walk models described in Section 2.1 can be implemented as ABMs). Further rules can then be added to describe the cell cycle and the cell interactions with its neighbours. In their simpler incarnations, ABMs contain explicit rules deciding each 'agent's' evolution, whereas further sophistication is achieved by deriving parameters that enter those rules from the solution of a system of differential equations. The idea of agent-based modelling is to learn something about the whole system (i.e., macroscale and collective behaviours) from various assumptions about the individual (or microscale) behaviour.

One of the key advantages of ABMs is their vast flexibility; it is simple to add more agents, change the governing rules and add or remove layers of complexity in the cell behaviour. They have been used to predict the emergent behaviour of cell interactions [100], model self-organisation within colonies [101] and describe stem cell driven tumour growth [102]. A disadvantage of the ABMs is their computational cost which increases rapidly with the number of the 'agents' and the sophistication of rules that govern their behaviour. It is also difficult to develop simplified analytical models in order to interpret and generalise the results of ABM simulations.

## 4.1 ABMs for cell migration

ABMs have been developed to incorporate the coordinated migration and collective behaviours of stem cells within colonies, but the challenges remain to fully capture the experimental behaviours, especially collective aspects and cell migration in three dimensions. These agent-based migration models are often combined with models of colony growth and proliferation [9, 103] (as discussed in Section 3.2).

hPSCs show coordinated intra-colony movements which cease upon differentiation [104]. Cell movement speed varies within colonies, with higher average speed at the periphery and lower in the central region [62]. Recently, a two-dimensional (monolayer) individual-based stochastic model was developed of cell migration, cell-cell connections and cell-substrate connections which captures well these experimental observations [9]. The model is formulated in terms of the energies associated with the cell-cell and cell-substrate connections. Any energy released by breaking and forming such connections drives cell migration in one of the eight directions on a square lattice along the lattice sides and diagonals. The direction of movement is determined at random according to a probability related to the cell's energy and a spatial weighting which favours the sideways rather than diagonal displacements (as described in [105]). Cell proliferation and quiescence are also included. This model suggests that cell division is a leading factor in the increased mobility at the colony edges. This is a promising approach for the studies of complex behaviours of iPSCs and planning of bio-processing experiments.

Models of cell cultures in three dimensions are developing actively as they are required to achieve a more realistic understanding of *in-vivo* behaviours of hPSCs [106] and *in-vitro* engineering of tissues on three-dimensional scaffolds [107]. Recently developed models provide a good starting point for such simulations, such as the open-source PhysiCell model originally devised for cancer cells but transferable to other cell types including iPSCs [36]. The model implements cell movement by defining a persistence time, a migration speed and a migration bias, allowing for a range of cell motions from purely random to deterministic. The model also includes mechanical interactions between neighbouring cells. Figure 13 shows an example of the outcome of such simulations of a hanging drop spheroid with deterministic and stochastic necrosis.

*Figure 13 - Hanging drop spheroid simulations with deterministic necrosis (left) and stochastic necrosis (right). Different cell states are colour coded. These simulations predict the emergence of a necrotic core microstructure arising from cell-scale mechanical interactions (adapted from Ref. [36]).*

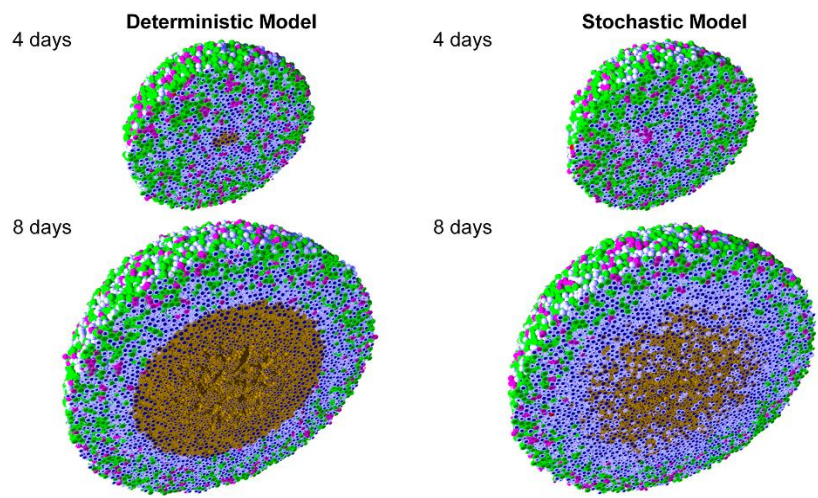

Modelling cell movement on a three-dimensional discrete lattice is widely used, e.g., for modelling the mesenchymal stem cell tissue differentiation [108] and tumour growth driven by cancer stem cells [102]. Some models allow many lattice nodes per cell representing a cell as a spatially extended object (as in the Potts model) [109]. There is also a range of agent-based continuous models where cell movement is not restricted to a grid but occurs continuously in any direction as illustrated in two dimensions in Figure 3 [110, 111]. Here the movement is described in terms of forces or potentials, with positions obtained from differential equations of motion for each cell. In centre-based models, each cell is represented by a simple geometrical object, such as a circle, whereas vertex models describe a cell as a number of connected nodes [112].

Another application of the ABM focusses on the cells' changing morphology. For example, a model for mesenchymal stem cells [103] includes the random formation, elongation and retraction of pseudopodia resulting in forces that lead to cell movement. This model can reproduce the spatio-temporal organisation of cells at a quantitative level and emphasises the importance of cell-cell interactions in tissue formation. In its published form, this model shows stronger inertial and accelerated movements than experimental results [39]. How much of this discrepancy is due to differences in cell type and culturing conditions needs to be investigated further to clarify the model's applicability to different experiments.

## 4.2 Colony proliferation

The ABMs discussed above are implemented on a fixed lattice, so that the 'agent' positions change in jumps from one lattice node to another. An alternative is to allow for an irregular shape and variable size of a region allocated to each 'agent' (cell). In one implementation of this approach, each cell is represented as a vertex in a graph whose edges are straight lines joining each cell with its nearest neighbours, as shown with dashed lines in Figure 14. To define the neighbourhood of each cell, a perpendicular line is drawn through the middle of each edge to form a closed polygon with a cell inside. Such a division of the plane into polygonal cells (Voronoi cells) is called the Voronoi tessellation. A Voronoi cell consists of all points that are closer to its 'agent' than to any other. Similar procedure in the three-dimensional space results in the space partition into convex polyhedrons. The Voronoi cells are not uniform in shape and have a variable number of sides depending on the number density of the cells. The tessellation can also be constructed to represent experimental images using the geometrical cell centroids or their nucleus positions, as shown in Figure 14 [33].

*Figure 14 – The Voronoi diagram (with edges shown black solid) illustrating how colony area is divided into tessellated cells around the 'agents' (blue circles) with the nearest neighbours joined by dashed lines . Right: The Voronoi tessellation obtained from the centroid positions of cells in an experimental microscopic image [33].*

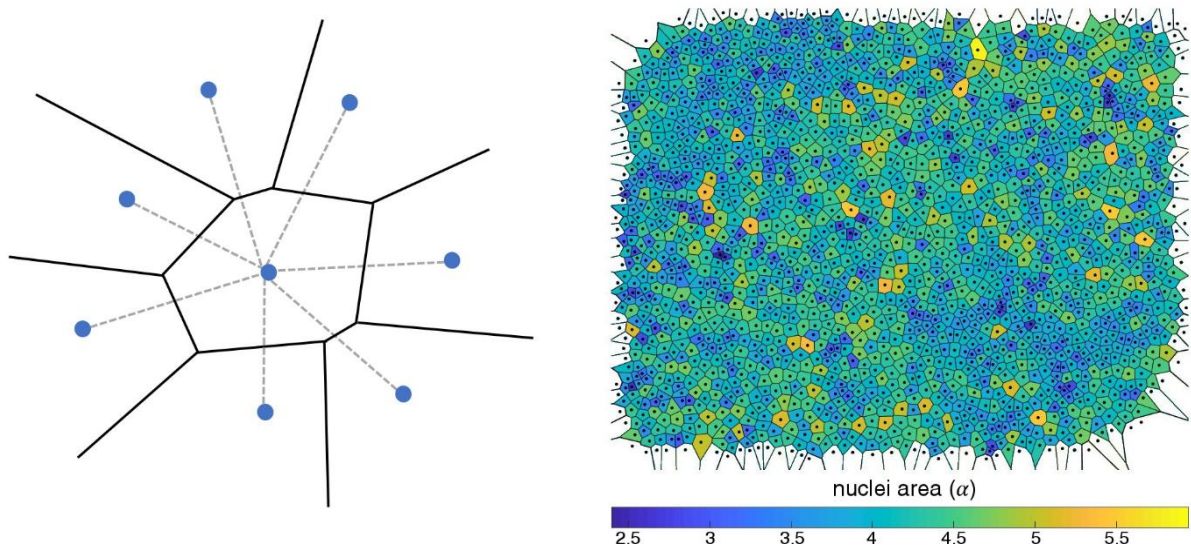

Voronoi tessellation has been used to model monolayer adult stem cells in intestinal crypts [113, 114] and is now being transferred to hPSCs. The model uses an agent-based approximation in which each cell is represented as a Voronoi tessellation of the space [113, 115]. The domain grows according to the inter-cellular pressure that builds up due to mitotic divisions in the colony. The cell-cell interactions are described by a force derived from an elastic potential given for a cell labelled with a number $i$ by

$$V(\mathbf{r}_i, t) = \frac{k_v}{2} [\alpha_i(t) - \overline{\alpha_0}(t)]^2 + \frac{k_c}{2} [\mathbf{r}_i(t) - \mathbf{r}_{0i}(t)]^2, \qquad (14)$$

where $k_v$ and $k_c$ are the elastic constants, $\alpha_i$ the area of the $i$'th cell, $\bar{\alpha}_0$ is the equilibrium cell area and $\mathbf{r}_i$ is the initial positions of the $i$'th cell which does not necessarily coincide with its centroids position $\mathbf{r}_{0i}$. The first term in the right-hand side of Eq. (14) reflects the tendency for the cells to have similar sizes, whereas the second term controls the cell shape. The gradient of the potential combined with a drag force gives the total force acting on each cell.

Figure 15 shows a simulated colony undergoing a cell division. Cells in the middle of the colony experience a higher pressure and show mitotic arrest, i.e. they do not divide.

*Figure 15 - Voronoi tessellation to simulate a proliferating hPSC colony. $N_c$ is the number of cells. The cells divide and give rise to two daughter cells, see highlighted cells outlined in yellow. The colour bar shows the elastic field in Eq. (14).*

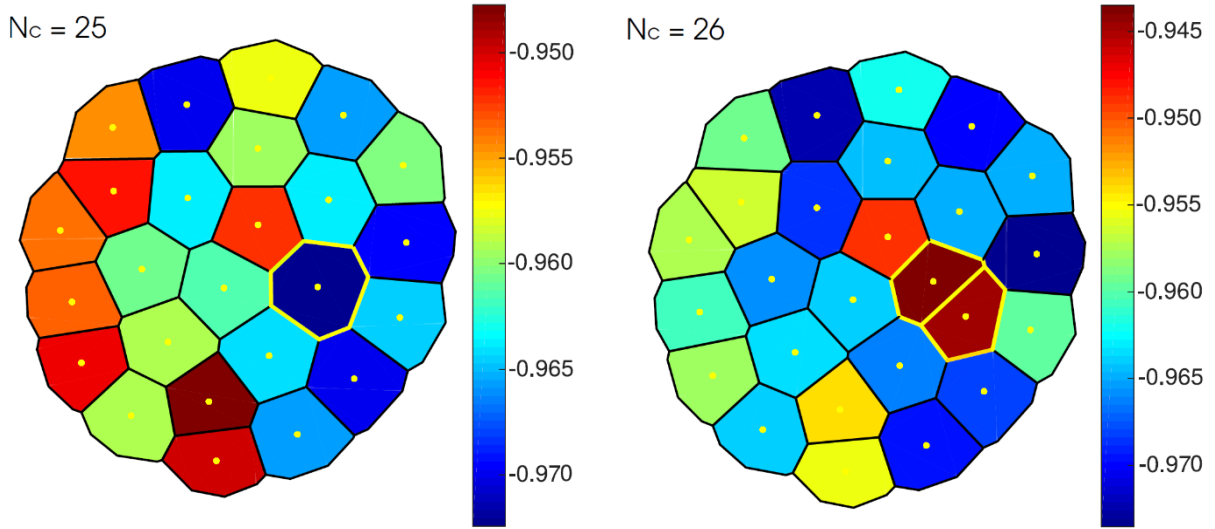

Spatially modelling each individual cell in a colony in this way raises an important question about the physical process involved in cell division: how does the colony rearrange to make space for new cells? In Voronoi tessellation models [113, 115] the cell rearrangement is controlled by the potential forces from the neighbouring cells and the crypt wall. In most square- or hexagonal-lattice models, one daughter cell is placed in the same position as the mother cell while the other is put into a neighbouring position, chosen at random [116], isotropic mitosis. If there is no free position available next to the dividing cell, the neighbouring cells are re-arranged into the free spaces available stochastically until there is a free space next to the dividing cell [9] or, if this is impossible, mitosis is suppressed (quiescence) [108, 117]. Such prescriptions are rather arbitrary and dedicated experimental time-lapse image data are needed to inform such models.

The cell proliferation in a colony is sensitive to various environmental factors. There is evidence that dense cell packing reduces cell proliferation [118], which has been captured in a model showing preferential cell division at the colony edge [9]. A self-organisation of cells has also been observed, where the newly divided (smallest) cells cluster together in patches, separated from larger cells at the final stages of the cell cycle [33]. This segregation by cell size allows neighbouring cells to move and swap places as the colony grows and could directly influence cell-to-cell interactions and community effects [33].

The PhysiCell model [36] described above allocates each cell a volume which varies with the cell cycle. The daughter cells have half the volume of their parent and are placed next to their parent cell position. A combination of this model and hPSC-specific spatial models is a promising approach to modelling the structure of three-dimensional colonies.

Spatial models of hPSCs become increasingly complex with colony size, and it is difficult to incorporate collective migratory effects. The question of how colonies rearrange upon cell divisions is pivotal in this approach and it requires deeper experimental investigation. These models have already advanced our understanding of the growth of cancer tumours [119] and wound healing [120].

# 5 iPSC specific models

*Synopsis: In this section, we review recent models describing the iPSC reprogramming. These models involve a variety of mathematical techniques and are unavoidably more complex than those previously discussed. Despite this, we aim to give the reader an overview of this type of models and highlight their usefulness for developing experiments.*

Although most of the models discussed so far have been constructed using hPSC data, their mathematical framework is adaptable to iPSCs. The advantage of many mathematical models is their transferable nature and flexibility which facilitate their adaptation to various experiments and data. In this section, we will discuss the recent advances in iPSC-specific modelling and the current challenges and successes in capturing iPSC behaviours and reprogramming.

## 5.1 iPSC reprogramming models

Mathematical models similar to those discussed in the previous sections are used to describe the reprogramming of somatic cells into iPSCs, which remains a low-yield process in the laboratory [121]. A combination of experimental effort and computational modelling has shown that iPSC reprogramming is a continuous stochastic process and so most mathematical models in this area are probabilistic [8]. Detailed reviews of the cell fate and reprogramming models can be found in Refs [122] and [74].

The modelling of the evolving cell state was first facilitated by Waddington's analogy [123] (illustrated in Figure 16) of a pluripotent cell as a ball positioned at the top of a hill: as the cell progressively loses its differentiation potential, it behaves similarly to a ball rolling downhill into a valley which represents what was then thought to be an irreversibly differentiated state. Although the Waddington landscape provides an intuitive representation, it lacks rigorous quantification and, in reality, the number of potential barriers to fate transitions are numerous. Some models simplify the transition between the two states (pluripotent and differentiated or somatic) to a one-step process [8] while others attempt to capture the transition through multiple stages [124], illustrated in Figure 16. Since the discovery of iPSCs (the equivalent of moving back up the hill in the Waddington landscape), experimental results have suggested a dynamical systems approach to modelling cell reprogramming [15] (a description of the evolution of a physical or biological system in terms of a system of ordinary differential equations).

*Figure 16 – Left: An illustration of Waddington's epigenetic landscape with the pluripotent state represented as a high-potential state and the differentiation described as a reduction in the potential associated with descent from a potential hill. Right: The transition process can be described as either a one-step process with only two states or as a gradual transition through many states or phases of pluripotency.*

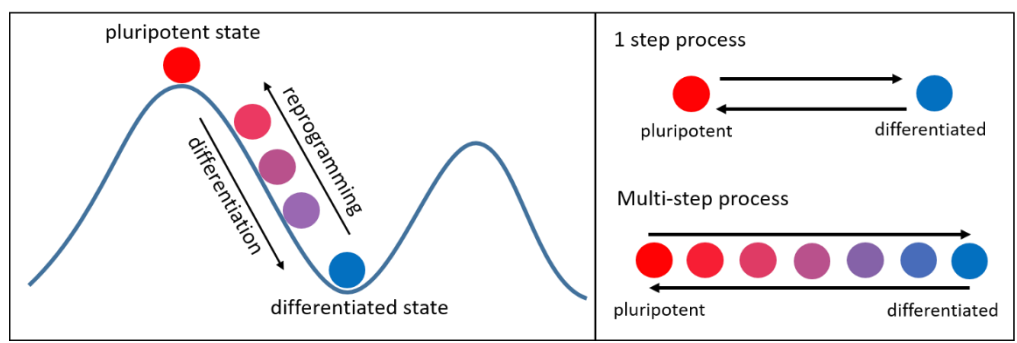

The pivotal study which suggested that the iPSC reprogramming is a continuous stochastic process did so through the experimental observation that almost any cell can be reprogrammed given a sufficient time (which is variable between cells) [8]. These results can be reproduced with one-step mathematical model. The model assumes that the reprogramming time is a random variable with the cumulative probability distribution $P = 1 - e^{-kt}$ where $k$ is a constant and $t$ is time normalized to the duration of the cell cycle. By comparing the model with the experimental data, the value of the transition rate $k$ has been found for various experimental conditions. It appears that the number of cell divisions is a key parameter driving epigenetic reprogramming.

Although the model broadly captures experimental data, it performs relatively poorly at early times suggesting that the transition rate $k$ can be time-dependent. This question was explored using a network model to suggest that the reprogramming is a stochastic process at early times but deterministic at later stages [125]. Thus, either two different models are required to capture the two phases, or a more complex model should incorporate both using a time-dependent degree of stochasticity.

Other models directly consider the genetic network involved in the reprogramming, and some include the epigenetic network [126]. A recent model has described cell types as a set of hierarchically related dynamical attractors representing cell cycles [22]. The model is streamlined to describe the gene expression levels of cell types across their cell cycles to simplify the mathematics. The evolution of the expression level of the gene $i$ due to binary gene interactions (with time $t$ measured in the units of the cell cycle time) is given by

$$n_i(t+1) = \Theta\left[\sum_j J_{ij} n_j(t) - \theta - T\zeta_i(t)\right], \tag{15}$$

where $n_i$ is the expression level of the gene $i$, $J_{ij}$ represents the interaction between the genes $i$ and $j$, $\theta$ is a threshold for neuron firing, $T$ a parameter that specifies the magnitude of the random noise $\zeta_i$, and $\Theta$ is the Heaviside step function $\Theta(x) = 0$ for $x \leq 0$ and $\Theta(x) = 1$ for $x > 0$. This model is a two-level hierarchy with stem cells at the top, $n_i = 1$, and the daughter cells at the bottom, $n_i = 0$. The gene expression levels are considered to be random quantities with the probability density

$$p(\eta^{\rho\mu}) = \begin{cases} a^{\rho\mu} & \text{for } \eta^{\rho\mu} = 1, \\ 1 - a^{\rho\mu} & \text{for } \eta^{\rho\mu} = 0, \end{cases} \tag{16}$$

where $\eta^{\rho\mu}$ is the expression of the gene during the $\rho$ phase of the cell cycle of daughter cell $\mu$ and $a^{\rho\mu}$ is the mean $\mu$-daughter type gene expression during the $\rho$ phase. Using the theory of neural networks to describe the interactions between the gene expression levels results in two equations for the dynamics which can be solved numerically. This model has led to the identification of two mechanisms for reprogramming in a two-level hierarchy: cycle-specific perturbations and a noise-induced switching [22]. Each of these reprogramming scenarios leads to specific patterns of the reprogramming dynamics that are broadly in line with experimental findings. The model shows it takes multiple cell cycle generations to become reprogrammed and that a finite fraction of gene expression levels need to be perturbed in order to reprogram a cell [22].

The complexity of such models unavoidably leads to a large number of model parameters and many assumptions, and not all of them can be justified or verified experimentally. Therefore, an agreement with a limited set of experiments does not necessarily indicate the relevance of each part of such a model. This problem is common to any kind of mathematical model and prompts one to prefer the most simple but effective models that rely on explicitly justifiable (rather than just plausible) assumptions and a broad range of experimental results.

Another aspect of cell evolution is the birth and death of cells, which can be treated in the framework of a two-state system where transitions between the states are modelled as a continuous-time Markov process (i.e., each transition is statistically independent of any earlier states) (see Refs [127] and [128] for further mathematical details and biological applications). The model of Ref. [34] uses such a process to explore the reprogramming of somatic cells. It allows for a probabilistic reprogramming rate consistent with experimental results. The model considers two cell types (somatic cells and iPSCs), uses the logistic growth model reviewed in Section 3.1 to describe proliferation, and introduces transition probabilities to represent the chance of a cell changing between the two states. The model reveals two different modes of cellular reprogramming dynamics: PTF expression alone leads to heterogeneous reprogramming while PTFs plus certain other factors homogenise the dynamics.

A wide variety of mathematical methods are currently being used to model iPSC reprogramming, not limited to the most common approaches of dynamical systems and stochastic processes. The complexity of the system and the lack of appropriate experimental data that could be used to constrain such models appear to be the key challenges in the model development, verification and testing. Nevertheless, mathematical modelling has already revealed interesting behaviours in the reprogramming dynamics and a quantitative understanding of cell state transitions will facilitate the formulation of improved reprogramming strategies and guide the optimisation of reprogramming experiments. An even deeper connection between mathematical and experimental efforts appears to be necessary. Dedicated experiments are required to constrain the models and limit the potentially unbounded diversity of mathematical approaches. On the other hand, a hierarchy of simple mathematical models aimed at explaining limited but significant aspects of progressively complex behaviours observed in the experiment may prove to be more successful in deepening our understanding than a single complicated model. Both sides of such a comprehensive approach to mathematical modelling must rely on a closer collaboration of the experimentalists and theoreticians.

# 6 Discussion and prospects

Mathematical models of cell behaviour are essential in developing non-invasive predictive tools of stem cell biology. Their purposes include the identification of inherent and universal systematic behaviours, determination of the key factors that control the cell behaviour, and quantitative comparison of experimental results obtained under different biological conditions.

The complexity of stem cell and iPSC systems that involve a plethora of chemical, mechanical and biological effects that act at a range of spatial and temporal scales aggravates the modelling and understanding of the intra-cellular, inter-cellular and colony-scale behaviours. A systematic approach to the modelling is essential, and models isolating a limited range of the key properties have often been more successful and useful in interpreting experimental results [35]. For example, focussed migration models (Section 2) have led to a greater understanding of the behaviour of isolated cells [16, 24, 25] and the movement of cells within colonies [9, 103]. The population models for colony proliferation (see Section 3.2) have been advanced, e.g., to include cell divisions and deaths, providing a distinct computational advantage over more complex spatio-temporal models. Such models have been used to investigate the impact of colony expansion on clonality [7], cell regeneration within intestinal crypts [113, 114] and tumour growth [119]. On the other hand, the complexity of the real systems, with large numbers of diverse factors that affect their evolution, makes them amenable to well-developed probabilistic approaches.

Many current efforts focus on modelling cell pluripotency and cell fate, as applications of hPSCs require greater control over pluripotency and differentiation trajectories. The stochastic nature of pluripotency at the single cell level [13], along with recent studies of the fluctuations of PTFs throughout colonies [129] and spatial patterning of differentiation [79, 80] are being used to inform the development of models of pluripotency and cell fate.

The focus for iPSC modelling is to capture the reprogramming process. Experimental results suggest a dynamical systems approach may be appropriate [8] and current models often make use of birth-death processes [34], cell cycle dynamical attractors [22] and modelling of genetic networks [22, 126]. These models are promising but more is required to establish consistent models of the reprogramming. Further experiments are essential to justify, compare and validate such models.

Many other general behaviours of iPSCs, such as cell migration and colony proliferation, have yet to be fully explored using mathematical models. The existing experience with hPSC modelling is clearly relevant but more experimental work is needed to quantify the similarities and differences between the iPSCs and other cell types.

The comparison of models with experiments represents a separate and important aspect of the modelling. Model refinement should be based on a two-way interaction with experiments; model parameters need to be better informed by experimental results. Mathematical models should be used to influence experimental design to achieve a deeper interpretation of the experiments, extending beyond relatively simple descriptive presentations. Models of this kind have already helped provide an insight into tissue formation, wound healing, tumour growth and the reprogramming of iPSCs and will no doubt continue to do so.

We emphasise that this interdisciplinary approach is an exciting and rewarding method for deepening our understanding of stem cell systems and encourage a continued discussion between biologists and mathematicians working towards a common goal.

***Acknowledgements:*** *IN thanks the grant from the Russian Government Program for the recruitment of the leading scientists into Russian Institution of Higher Education.14.w03.31.0029 and RFFI project grant number 20-015-0060.*

# Bibliography

1. R. Barnes and C. Lehman, *Modeling of bovine spongiform encephalopathy in a two-species feedback loop.* Epidemics, 2013. **5**(2): p. 85-91.
2. M. Al-Zoughool, et al., *Mathematical models for estimating the risks of bovine spongiform encephalopathy (bse).* J. Toxicol. Environ. Health. B Crit. Rev., 2015. **18**(2): p. 71-104.
3. S. M. Kissler, et al., *Geographic transmission hubs of the 2009 influenza pandemic in the united states.* Epidemics, 2019. **26**: p. 86-94.
4. M. R. Servedio, et al., *Not just a theory—the utility of mathematical models in evolutionary biology.* PLoS Biol., 2014. **12**(12): p. e1002017.
5. T. Hillen and K. J. Painter, *A user's guide to pde models for chemotaxis.* J. Math. Biol., 2009. **58**(1-2): p. 183-217.
6. P. M. Altrock, L. L. Liu, and F. Michor, *The mathematics of cancer: Integrating quantitative models.* Nat. Rev. Cancer, 2015. **15**(12): p. 730-45.
7. L. E. Wadkin, et al., *Seeding hescs to achieve optimal colony clonality.* Sci. Rep., 2019. **9**(1): p. 15299.
8. J. Hanna, et al., *Direct cell reprogramming is a stochastic process amenable to acceleration.* Nature, 2009. **462**(7273): p. 595-601.
9. T. N. T. Nguyen, K. Sasaki, and M. Kino-oka, *Elucidation of human induced pluripotent stem cell behaviors in colonies based on a kinetic model.* J. Biosci. Bioeng., 2019. **127**(5): p. 625-632.
10. L. E. Wadkin, et al., *The recent advances in the mathematical modelling of human pluripotent stem cells.* SN Applied Sciences, 2020. **2**(2): p. 276.
11. J. D. Murray, *Mathematical biology.* 3 ed. Interdisciplinary applied mathematics. 2002: Springer-Verlag New York. XXIII, 551.
12. J. D. Murray, *Mathematical biology ii.* Interdisciplinary applied mathematics. 2003: Springer-Verlag New York. XXV, 814.
13. B. D. MacArthur and I. R. Lemischka, *Statistical mechanics of pluripotency.* Cell, 2013. **154**(3): p. 484-489.
14. M. Li and J. C. Izpisua Belmonte, *Deconstructing the pluripotency gene regulatory network.* Nat. Cell Biol., 2018. **20**(4): p. 382-392.
15. S. Huang, et al., *Cell fates as high-dimensional attractor states of a complex gene regulatory network.* PhRvL, 2005. **94**(12): p. 128701.
16. L. E. Wadkin, et al., *Dynamics of single human embryonic stem cells and their pairs: A quantitative analysis.* Sci. Rep., 2017. **7**(1): p. 570.
17. M. Tewary, et al., *A stepwise model of reaction-diffusion and positional information governs self-organized human peri-gastrulation-like patterning.* Development, 2017. **144**(23): p. 4298.
18. J. E. Till, E. A. McCulloch, and L. Siminovitch, *A stochastic model of stem cell proliferation, based on the growth of spleen colony-forming cells.* Proc. Natl. Acad. Sci. U. S. A., 1964. **51**(1): p. 29-36.
19. P. Pir and N. Le Novere, *Mathematical models of pluripotent stem cells: At the dawn of predictive regenerative medicine.* Methods Mol. Biol., 2016. **1386**: p. 331-50.
20. V. Olariu and C. Peterson, *Kinetic models of hematopoietic differentiation.* WIREs Systems Biology and Medicine, 2019. **11**(1): p. e1424.
21. H. Xu, et al., *Construction and validation of a regulatory network for pluripotency and self-renewal of mouse embryonic stem cells.* PLoS Comput. Biol., 2014. **10**(8): p. e1003777.
22. R. Hannam, A. Annibale, and R. Kühn, *Cell reprogramming modelled as transitions in a hierarchy of cell cycles.* J. Phys. A, 2017. **50**(42): p. 425601.

23. A. L. MacLean, P. D. W. Kirk, and M. P. H. Stumpf, *Cellular population dynamics control the robustness of the stem cell niche.* Biology Open, 2015. **4**(11): p. 1420.
24. L. Li, et al., *Individual cell movement, asymmetric colony expansion, rho-associated kinase, and e-cadherin impact the clonogenicity of human embryonic stem cells.* Biophys. J., 2010. **98**(11): p. 2442-2451.
25. L. E. Wadkin, et al., *Correlated random walks of human embryonic stem cells in vitro.* Phys. Biol., 2018. **15**(5): p. 056006.
26. J. M. Raimond, et al., *Collective absorption of blackbody radiation by rydberg atoms in a cavity: An experiment on bose statistics and brownian motion.* PhRvL, 1982. **49**(2): p. 117-120.
27. M. de Jager, et al., *How superdiffusion gets arrested: Ecological encounters explain shift from lévy to brownian movement.* Proceedings of the Royal Society B: Biological Sciences, 2014. **281**(1774): p. 20132605.
28. M. Batty, *Agent-based pedestrian modelling.* Advanced spatial analysis: The CASA book of GIS, 2003. **81**: p. 81-106.
29. A. J. Loosley, et al., *Describing directional cell migration with a characteristic directionality time.* PLoS One, 2015. **10**(5): p. e0127425.
30. I. Barbaric, et al., *Time-lapse analysis of human embryonic stem cells reveals multiple bottlenecks restricting colony formation and their relief upon culture adaptation.* Stem Cell Reports, 2014. **3**(1): p. 142-155.
31. C. W. Gardiner, *Handbook of stochastic methods*. Vol. 3. 1985: springer Berlin.
32. B. B. Mandelbrot and J. W. Van Ness, *Fractional brownian motions, fractional noises and applications.* SIAMR, 1968. **10**(4): p. 422-437.
33. S. Orozco-Fuentes, et al., *Quantification of the morphological characteristics of hesc colonies.* Sci. Rep., 2019. **9**(1): p. 17569.
34. L. L. Liu, et al., *Probabilistic modeling of reprogramming to induced pluripotent stem cells.* Cell Rep., 2016. **17**(12): p. 3395-3406.
35. M. Tewary, N. Shakiba, and P. W. Zandstra, *Stem cell bioengineering: Building from stem cell biology.* Nature Reviews Genetics, 2018. **19**(10): p. 595-614.
36. A. Ghaffarizadeh, et al., *Physicell: An open source physics-based cell simulator for 3-d multicellular systems.* PLoS Comp. Biol., 2018. **14**(2): p. e1005991.
37. E. A. Codling, M. J. Plank, and S. Benhamou, *Random walk models in biology.* Journal of The Royal Society Interface, 2008. **5**(25): p. 813-834.
38. P. Turchin, *Quantitative analysis of movement: Measuring and modeling population redistribution in animals and plants*. 1998: Sinauer.
39. P. Dieterich, et al., *Anomalous dynamics of cell migration.* Proceedings of the National Academy of Sciences, 2008. **105**(2): p. 459.
40. P.-H. Wu, et al., *Three-dimensional cell migration does not follow a random walk.* Proceedings of the National Academy of Sciences, 2014. **111**(11): p. 3949.
41. J.-P. Bouchaud and A. Georges, *Anomalous diffusion in disordered media: Statistical mechanisms, models and physical applications.* PhR, 1990. **195**(4-5): p. 127-293.
42. W. M. Getz and D. Saltz, *A framework for generating and analyzing movement paths on ecological landscapes.* Proceedings of the National Academy of Sciences, 2008. **105**(49): p. 19066-19071.
43. S. Benhamou, *How to reliably estimate the tortuosity of an animal's path:: Straightness, sinuosity, or fractal dimension?* J. Theor. Biol., 2004. **229**(2): p. 209-220.
44. F. Bartumeus, et al., *Animal search strategies: A quantitative random-walk analysis.* Ecology, 2005. **86**(11): p. 3078-3087.
45. G. F. Lawler, *Random walk and the heat equation*. Vol. 55. 2010: American Mathematical Soc.

46. R. B. Dickinson, S. Guido, and R. T. Tranquillo, *Biased cell migration of fibroblasts exhibiting contact guidance in oriented collagen gels.* Ann. Biomed. Eng., 1994. **22**(4): p. 342-356.
47. R. B. Dickinson, *A generalized transport model for biased cell migration in an anisotropic environment.* J. Math. Biol., 2000. **40**(2): p. 97-135.
48. W. Alt, *Biased random walk models for chemotaxis and related diffusion approximations.* J. Math. Biol., 1980. **9**(2): p. 147-177.
49. F. A. C. C. Chalub, et al. *Kinetic models for chemotaxis and their drift-diffusion limits.* in *Nonlinear Differential Equation Models.* 2004. Vienna: Springer Vienna.
50. A. Okubo and S. A. Levin, *Diffusion and ecological problems: Modern perspectives.* Vol. 14. 2013: Springer Science & Business Media.
51. M. H. Gail and C. W. Boone, *The locomotion of mouse fibroblasts in tissue culture.* Biophys. J., 1970. **10**(10): p. 980-93.
52. R. L. Hall, *Amoeboid movement as a correlated walk.* J. Math. Biol., 1977. **4**(4): p. 327-35.
53. A. A. Potdar, et al., *Human mammary epithelial cells exhibit a bimodal correlated random walk pattern.* PLoS One, 2010. **5**(3): p. e9636.
54. P. M. Kareiva and N. Shigesada, *Analyzing insect movement as a correlated random walk.* Oecologia, 1983. **56**(2): p. 234-238.
55. C. M. Bergman, J. A. Schaefer, and S. N. Luttich, *Caribou movement as a correlated random walk.* Oecologia, 2000. **123**(3): p. 364-374.
56. D. S. Johnson, et al., *Continuous-time correlated random walk model for animal telemetry data.* Ecology, 2008. **89**(5): p. 1208-1215.
57. E. Batschelet, *Circular statistics in biology.* 1981: Academic Press.
58. P. Berens, *Circstat: A matlab toolbox for circular statistics.* 2009, 2009. **31**(10): p. 21.
59. A. Pewsey, M. Neuhäuser, and G. D. Ruxton, *Circular statistics in r.* 2013: OUP Oxford.
60. I. Glauche, et al., *Stem cell clonality -- theoretical concepts, experimental techniques, and clinical challenges.* Blood Cells Mol. Dis., 2013. **50**(4): p. 232-40.
61. N. Heins, et al., *Clonal derivation and characterization of human embryonic stem cell lines.* J. Biotechnol., 2006. **122**(4): p. 511-520.
62. E. Shuzui, M.-H. Kim, and M. Kino-oka, *Anomalous cell migration triggers a switch to deviation from the undifferentiated state in colonies of human induced pluripotent stems on feeder layers.* J. Biosci. Bioeng., 2019. **127**(2): p. 246-255.
63. A. L. MacLean, C. Lo Celso, and M. P. Stumpf, *Concise review: Stem cell population biology: Insights from hematopoiesis.* Stem Cells, 2017. **35**(1): p. 80-88.
64. F. Michor, *Mathematical models of cancer stem cells.* J. Clin. Oncol., 2008. **26**(17): p. 2854-2861.
65. M. Loeffler and H. E. Wichmann, *A comprehensive mathematical model of stem cell proliferation which reproduces most of the published experimental results.* Cell Tissue Kinet., 1980. **13**(5): p. 543-61.
66. B. M. Deasy, et al., *Modeling stem cell population growth: Incorporating terms for proliferative heterogeneity.* Stem Cells, 2003. **21**(5): p. 536-545.
67. J. L. Sherley, P. B. Stadler, and J. S. Stadler, *A quantitative method for the analysis of mammalian cell proliferation in culture in terms of dividing and non-dividing cells.* Cell Prolif., 1995. **28**(3): p. 137-144.
68. M. A. Tabatabai, et al., *Mathematical modeling of stem cell proliferation.* Med Biol Eng Comput, 2011. **49**(3): p. 253-262.
69. M. Tabatabai, D. K. Williams, and Z. Bursac, *Hyperbolastic growth models: Theory and application.* Theoretical Biology and Medical Modelling, 2005. **2**(1): p. 14.

70. G. Chen, et al., *Actin-myosin contractility is responsible for the reduced viability of dissociated human embryonic stem cells.* Cell stem cell, 2010. **7**(2): p. 240-248.
71. D. Moogk, et al., *Human esc colony formation is dependent on interplay between self-renewing hescs and unique precursors responsible for niche generation.* Cytometry Part A, 2010. **77A**(4): p. 321-327.
72. M. Amit, et al., *Clonally derived human embryonic stem cell lines maintain pluripotency and proliferative potential for prolonged periods of culture.* Dev. Biol., 2000. **227**(2): p. 271-278.
73. O. Symmons and A. Raj, *What's luck got to do with it: Single cells, multiple fates, and biological nondeterminism.* Mol. Cell, 2016. **62**(5): p. 788-802.
74. M. Herberg and I. Roeder, *Computational modelling of embryonic stem-cell fate control.* Development, 2015. **142**(13): p. 2250.
75. B. D. MacArthur, A. Ma'ayan, and I. R. Lemischka, *Systems biology of stem cell fate and cellular reprogramming.* Nature Reviews Molecular Cell Biology, 2009. **10**(10): p. 672-681.
76. S. Viswanathan and P. W. Zandstra, *Towards predictive models of stem cell fate.* Cytotechnology, 2003. **41**(2-3): p. 75-92.
77. R. M. Kumar, et al., *Deconstructing transcriptional heterogeneity in pluripotent stem cells.* Nature, 2014. **516**(7529): p. 56-61.
78. J. Jang, et al., *Control over single-cell distribution of g1 lengths by wnt governs pluripotency.* PLoS Biol., 2019. **17**(9): p. e3000453.
79. A. Warmflash, et al., *A method to recapitulate early embryonic spatial patterning in human embryonic stem cells.* Nat. Methods, 2014. **11**(8): p. 847-854.
80. K. A. Rosowski, et al., *Edges of human embryonic stem cell colonies display distinct mechanical properties and differentiation potential.* Sci. Rep., 2015. **5**(1): p. 14218.
81. V. Chickarmane, et al., *Transcriptional dynamics of the embryonic stem cell switch.* PLoS Comp. Biol., 2006. **2**(9): p. e123.
82. V. Likhoshvai and A. Ratushny, *Generalized hill function method for modeling molecular processes.* J. Bioinf. Comput. Biol., 2007. **5**(02b): p. 521-531.
83. R. N. Gutenkunst, et al., *Universally sloppy parameter sensitivities in systems biology models.* PLoS Comp. Biol., 2007. **3**(10): p. e189.
84. I. Glauche, M. Herberg, and I. Roeder, *Nanog variability and pluripotency regulation of embryonic stem cells--insights from a mathematical model analysis.* PLoS One, 2010. **5**(6): p. e11238-e11238.
85. T. Kalmar, et al., *Regulated fluctuations in nanog expression mediate cell fate decisions in embryonic stem cells.* PLoS Biol., 2009. **7**(7).
86. Y. Luo, et al., *Cell signalling regulates dynamics of nanog distribution in embryonic stem cell populations.* Journal of The Royal Society Interface, 2013. **10**(78): p. 20120525.
87. V. Olariu, C. Lövkvist, and K. Sneppen, *Nanog, oct4 and tet1 interplay in establishing pluripotency.* Sci. Rep., 2016. **6**(1): p. 25438.
88. R. C. G. Smith, et al., *Nanog fluctuations in embryonic stem cells highlight the problem of measurement in cell biology.* Biophys. J., 2017. **112**(12): p. 2641-2652.
89. P. Yu, et al., *Nanog induced intermediate state in regulating stem cell differentiation and reprogramming.* BMC Syst. Biol., 2018. **12**(1): p. 22.
90. I. R. Akberdin, et al., *Pluripotency gene network dynamics: System views from parametric analysis.* PLoS One, 2018. **13**(3): p. e0194464-e0194464.
91. I. Chambers and A. Smith, *Self-renewal of teratocarcinoma and embryonic stem cells.* Oncogene, 2004. **23**(43): p. 7150-7160.

92. D. Auddya and B. J. Roth, *A mathematical description of a growing cell colony based on the mechanical bidomain model.* Journal of Physics D: Applied Physics, 2017. **50**(10): p. 105401.
93. C. S. Henriquez, *Simulating the electrical behavior of cardiac tissue using the bidomain model.* Crit. Rev. Biomed. Eng., 1993. **21**(1): p. 1-77.
94. W. M. Eby and N. Coleman, *Mathematical models in stem cell differentiation and fate predictability*, in *Regenerative medicine - from protocol to patient: 1. Biology of tissue regeneration*, G. Steinhoff, Editor. 2016, Springer International Publishing: Cham. p. 175-222.
95. S. R. K. Vedula, et al., *Collective cell migration: A mechanistic perspective.* Physiology, 2013. **28**(6): p. 370-379.
96. K. Muguruma, et al., *Self-organization of polarized cerebellar tissue in 3d culture of human pluripotent stem cells.* Cell Rep., 2015. **10**(4): p. 537-550.
97. S. Dekoninck and C. Blanpain, *Stem cell dynamics, migration and plasticity during wound healing.* Nat. Cell Biol., 2019. **21**(1): p. 18-24.
98. K. Rossington and T. Benson, *An agent-based model to predict fish collisions with tidal stream turbines.* Renewable Energy, 2019.
99. C. M. Henein and T. White. *Agent-based modelling of forces in crowds*. in *International Workshop on Multi-Agent Systems and Agent-Based Simulation*. 2004. Springer.
100. D. C. Walker, et al., *The epitheliome: Agent-based modelling of the social behaviour of cells.* BioSyst., 2004. **76**(1): p. 89-100.
101. M. d'Inverno and R. Saunders. *Agent-based modelling of stem cell self-organisation in a niche*. in *Engineering Self-Organising Systems*. 2005. Berlin, Heidelberg: Springer Berlin Heidelberg.
102. J. Poleszczuk, P. Macklin, and H. Enderling, *Agent-based modeling of cancer stem cell driven solid tumor growth.* Methods Mol. Biol., 2016. **1516**: p. 335-346.
103. M. Hoffmann, et al., *Spatial organization of mesenchymal stem cells in vitro—results from a new individual cell-based model with podia.* PLoS One, 2011. **6**(7): p. e21960.
104. Thomas A. Zangle, et al., *Quantification of biomass and cell motion in human pluripotent stem cell colonies.* Biophys. J., 2013. **105**(3): p. 593-601.
105. M. Kino-Oka, et al., *Valuation of growth parameters in monolayer keratinocyte cultures based on a two-dimensional cell placement model.* J. Biosci. Bioeng., 2000. **89**(3): p. 285-7.
106. Y. Lei, et al., *Developing defined and scalable 3d culture systems for culturing human pluripotent stem cells at high densities.* Cell. Mol. Bioeng., 2014. **7**(2): p. 172-183.
107. X. Yin, et al., *Engineering stem cell organoids.* Cell Stem Cell, 2016. **18**(1): p. 25-38.
108. H. Khayyeri, et al., *Corroboration of mechanobiological simulations of tissue differentiation in an in vivo bone chamber using a lattice-modeling approach.* J. Orth. Res., 2009. **27**(12): p. 1659-1666.
109. A. Szabó and R. M. H. Merks, *Cellular potts modeling of tumor growth, tumor invasion, and tumor evolution.* Front. Oncol., 2013. **3**: p. 87-87.
110. S. Adra, et al., *Development of a three dimensional multiscale computational model of the human epidermis.* PLoS One, 2010. **5**(1): p. e8511.
111. P. Van Liedekerke, A. Buttenschön, and D. Drasdo, *Chapter 14 - off-lattice agent-based models for cell and tumor growth: Numerical methods, implementation, and applications*, in *Numerical methods and advanced simulation in biomechanics and biological processes*, M. Cerrolaza, S.J. Shefelbine, and D. Garzón-Alvarado, Editors. 2018, Academic Press. p. 245-267.

112. P. Van Liedekerke, et al., *Simulating tissue mechanics with agent-based models: Concepts, perspectives and some novel results.* Computational Particle Mechanics, 2015. **2**(4): p. 401-444.
113. O.-F. Sirio and R. A. Barrio, *Modelling the dynamics of stem cells in colonic crypts.* The European Physical Journal Special Topics, 2017. **226**(3): p. 353-363.
114. F. A. Meineke, C. S. Potten, and M. Loeffler, *Cell migration and organization in the intestinal crypt using a lattice-free model.* Cell Prolif., 2001. **34**(4): p. 253-266.
115. R. A. Barrio, et al., *Cell patterns emerge from coupled chemical and physical fields with cell proliferation dynamics: The arabidopsis thaliana root as a study system.* PLoS Comp. Biol., 2013. **9**(5): p. e1003026.
116. D. Lehotzky and G. K. H. Zupanc, *Cellular automata modeling of stem-cell-driven development of tissue in the nervous system.* Dev. Neurobiol., 2019. **79**(5): p. 497-517.
117. M. A. Pérez and P. J. Prendergast, *Random-walk models of cell dispersal included in mechanobiological simulations of tissue differentiation.* J. Biomech., 2007. **40**(10): p. 2244-2253.
118. J. Wu, Y. Fan, and E. S. Tzanakakis, *Increased culture density is linked to decelerated proliferation, prolonged g1 phase, and enhanced propensity for differentiation of self-renewing human pluripotent stem cells.* Stem Cells Dev., 2014. **24**(7): p. 892-903.
119. Z. Wang, et al., *Simulating cancer growth with multiscale agent-based modeling.* Semin. Cancer Biol., 2015. **30**: p. 70-8.
120. D. Tartarini and E. Mele, *Adult stem cell therapies for wound healing: Biomaterials and computational models.* Frontiers in bioengineering and biotechnology, 2016. **3**: p. 206-206.
121. K. Takahashi, et al., *Induction of pluripotent stem cells from adult human fibroblasts by defined factors.* Cell, 2007. **131**(5): p. 861-872.
122. R. Morris, et al., *Mathematical approaches to modeling development and reprogramming.* Proceedings of the National Academy of Sciences, 2014. **111**(14): p. 5076.
123. C. H. Waddington, *The strategy of the genes. A discussion of some aspects of theoretical biology. With an appendix by h. Kacser.* 1957: London: George Allen & Unwin, Ltd. ix +-262 pp.
124. Y. Rais, et al., *Deterministic direct reprogramming of somatic cells to pluripotency.* Nature, 2013. **502**(7469): p. 65-70.
125. Y. Buganim, et al., *Single-cell expression analyses during cellular reprogramming reveal an early stochastic and a late hierarchic phase.* Cell, 2012. **150**(6): p. 1209-1222.
126. M. N. Artyomov, A. Meissner, and A. K. Chakraborty, *A model for genetic and epigenetic regulatory networks identifies rare pathways for transcription factor induced pluripotency.* PLoS Comp. Biol., 2010. **6**(5): p. e1000785.
127. M. Kijima, *Birth—death processes*, in *Markov processes for stochastic modeling*, M. Kijima, Editor. 1997, Springer US: Boston, MA. p. 243-293.
128. A. S. Novozhilov, G. P. Karev, and E. V. Koonin, *Biological applications of the theory of birth-and-death processes.* Brief. Bioinform., 2006. **7**(1): p. 70-85.
129. S. C. Wolff, et al., *Inheritance of oct4 predetermines fate choice in human embryonic stem cells.* Mol. Syst. Biol., 2018. **14**(9): p. e8140.